\documentclass{aa}  
\makeatletter
\renewcommand*\aa@pageof{, page \thepage{} of \pageref*{LastPage}}
\makeatother
\usepackage{graphicx}
\usepackage{txfonts}
\usepackage{multirow}
\usepackage{subcaption}
\usepackage[switch]{lineno}
\usepackage{hyperref}
\usepackage{cleveref}
\hypersetup{
    colorlinks=true,
    linkcolor=blue,
    filecolor=magenta,      
    urlcolor=cyan,
    citecolor=blue,
}

\usepackage{graphicx,psfig,epsf}  
\usepackage{times}
\usepackage{textcomp}
\usepackage{multirow, longtable}
\usepackage{amsmath}	% Advanced maths commands
\usepackage{amssymb}	% Extra maths symbols
\usepackage{gensymb}
\usepackage{colortbl}

\usepackage{array,etoolbox}
\preto\tabular{\setcounter{magicrownumbers}{0}}
\newcounter{magicrownumbers}
\def \rownumber{}

\def\apjl{ApJ}                % Astrophysical Journal, Letters
\def\apjs{ApJS}               % Astrophysical Journal, Supplement
\def\ao{Appl.~Opt.}           % Applied Optics
\def\aap{A\&A}                % Astronomy and Astrophysics
\def\src {NGC 4559 X7}
\def\ulx2 {NGC 4559 ULX-2}
\def\swift{{\it Swift/XRT}}
\def\nus{{\it NuSTAR}}
\def\chandra{{\it Chandra}}
\def\xmm{{\it XMM-Newton}}

\def\lum{erg s$^{-1}$}
\def\msun{M$_{\odot}$}

\usepackage[normalem]{ulem}

\begin{document}

\title{A new pulsating neutron star in the Ultraluminous X-ray source NGC 4559 X7?}
\titlerunning{A new PULX in NGC 4559 X7?}

\author{F. Pintore\inst{1}\thanks{E-mail: fabio.pintore@inaf.it}, C. Pinto\inst{1},  G. Rodriguez-Castillo\inst{1}, G.L. Israel\inst{2},  N. O. Pinciroli Vago\inst{2}\fnmsep\inst{10},  S. Motta$^{3}$,  F. Barra$^{1,12}$, D. J. Walton$^{4}$,  F. Fuerst$^{5}$,  P. Kosec$^{6}$,  C. Salvaggio$^{3}$,  M. Del Santo$^{1}$,  A. Wolter$^{3}$,  M. Middleton$^{7}$, A. D'A\`i$^{1}$,   E. Ambrosi$^{1}$,  L. Burderi$^{1,8}$,  M. Imbrogno$^{2,9}$,  R. Salvaterra$^{11}$, A. Robba$^{1,12}$ } 
\authorrunning{Pintore et al.}
\institute{INAF -- IASF Palermo, Via U. La Malfa 153, 90146 Palermo, Italy; \and INAF -- Osservatorio Astronomico di Roma, Via Frascati 33, I-00078, Monte Porzio Catone, Italy; \and  INAF -- Osservatorio Astronomico di Brera, via Brera 28, 20121 Milano, Italy; \and Centre for Astrophysics Research, University of Hertfordshire, College Lane, Hatfield AL10 9AB, UK;  \and European Space Agency (ESA), European Space Astronomy Centre (ESAC), Camino Bajo del Castillo s/n, 28692 Villanueva de la Ca\~nada, Madrid, Spain; \and  Center for Astrophysics | Harvard \& Smithsonian, Cambridge, MA, USA; \and School of Physics \& Astronomy, University of Southampton, Southampton, Southampton SO17 1BJ, UK; \and Dipartimento di Fisica, Universit\`a degli Studi di Cagliari, SP Monserrato-Sestu, KM 0.7, Monserrato, 09042, Italy; \and Dipartimento di Fisica, Università degli Studi di Roma ``Tor Vergata'', via della Ricerca Scientifica 1, I-00133 Rome, Italy; \and Department of Electronics, Information and Bioengineering, Politecnico di Milano, via G. Ponzio, 34, I-20133 Milan, Italy; \and INAF/IASF Milano, via Alfonso Corti 12, I-20133, Milano, Italy; \and Università degli Studi di Palermo, Dipartimento di Fisica e Chimica, Via Archirafi 36, 90123 Palermo, Italy; }

\date{Received ; accepted }

\abstract {{ Ultraluminous X-ray sources (ULX) are extragalactic objects with observed X-ray luminosities largely above the Eddington limit for a 10 M$_{\odot}$ black hole. It is nowadays believed that ULXs host super-Eddington accreting neutron stars or stellar mass black holes. However, the exact proportion of the two populations of compact objects is not yet known.} }
{{ We investigate the properties of the ULX NGC 4559 X7, which shows flux variability up to a factor of 5 on both long (months to years) and short (hours to days) timescales. A flaring activity was also observed during the source highest flux epochs. Flares are unpredictable, with different durations (but similar rising/decay times) and all flat-topped in flux. The latter suggests that, at the flare peaks, there is likely a common switch-off mechanism for the accretion onto the compact object.}}
{We analysed all the available \xmm\ and \swift\ observations in order to fully investigate the spectral and temporal evolution of X7, looking for short and long-term variability. We applied a Lomb-Scargle search to look for long-term periodicities. We also look for coherent signals through accelerated searches that included orbital corrections. We described the X7 spectral properties with two thermal components plus a cut-off powerlaw model. }
{We found three well defined spectral states. where the spectral variability is mainly driven by the two harder components, with the thermal one clearly following a correlation between its temperature and luminosity. In addition, a pulsed signal at 2.6~s--2.7~s was detected in two \xmm\ observations. The significance of these coherent signals is relatively weak but they are found in two different observations with the same parameter space for the orbital properties. If confirmed, it would imply a high spin-down of $10^{-9}$ s s$^{-1}$, which could be extreme amongst the known pulsating ULXs. X7 would become a new extragalactic ULX pulsar. }
{We discuss the spectral and temporal results of X7 in the context of super-Eddington accretion onto a stellar-mass compact object, in particular suggesting that the source might likely host a neutron star. }


\maketitle

\section{Introduction}
\label{introduction}

The number of ultraluminous X-ray sources (ULXs) counts about two thousands of objects \citep[e.g.][]{walton22}, characterized by high assumed isotropic X-ray luminosities well in excess of the Eddington limit for a 10 M$_{\odot}$ black hole (BH; for recent reviews see, e.g. \citealt{kaaret17,king23,pinto23}). 
Although some of them are still believed to host sub-Eddington accreting intermediate mass BHs ($10^2-10^5$ M$_{\odot}$; \citealt{colbert99}), most of the ULX population is likely composed by super-Eddington accreting stellar mass compact objects, i.e. BHs (M$_{BH}=5-80$ \msun) or neutron stars (NS). The number of confirmed NSs in the ULXs has grown significantly in the last decade, with the discovery of at least six (extragalactic) pulsating ULXs (PULXs), which is the smoking gun for a NS nature of the compact object in such sources (\citealt{bachetti14,israel16a}; \citealt{carpano18,israel16b}; \citealt{fuerst16,sathyaprakash19,rodriguez19}). Luminosities in PULXs can be very extreme, in some cases up to 1000 times their Eddington limit (\citealt{israel16a}). PULXs reside in binary systems with young and massive companion stars, where orbital periods ranges between 1--100 days and projected semi-major axis of a few-to-tens light seconds \citep[e.g.][]{bachetti22,belfiore24}.
Pulsations in PULXs are intermittent and characterised by a small pulsed fraction, which makes a challenge to increase the number of NSs in ULXs. However, it cannot be excluded that the NS population in ULXs is higher than currently known. On the basis of the PULX properties, indirect pieces of evidence \citep[e.g.][]{middleton17,king20} can be adopted to unveil candidate NSs in other ULXs: in particular, this evidence can be based on a switch between accretion and quiescent epochs during the ULX lifetime, believed to be due to the ignition of the propeller \citep[e.g.][]{tsygankov16,earnshaw19,song20,chashkina17, middleton23}; a hard spectrum \citep[e.g.][]{pintore17,koliopanos17,walton18,walton20,gurpide21}, likely associated to the emission from an accretion column at the NS polar caps; or to the detection of cyclotron absorption lines (as for M51 ULX-8; \citealt{brightman18,middleton19}).

In order to achieve luminosities as high as to $10^{41}$ \lum, a stellar-mass compact object needs to accrete matter well above the Eddington limit ($L_{\rm Edd} = 1.4 \times 10^{38} (M/M_{\odot})$ erg s$^{-1}$). At accretion rates $\dot{M} \gtrsim \dot{M}_{\rm Edd} = L_{\rm Edd} / (\eta c^2)$ ($\eta$ is the efficiency, $c$ the speed of light), radiation pressure is expected to thicken the accretion disc, which then exhibits thermal X-ray spectra peaking at a few keV with a high-energy cut-off around 10 keV. This is confirmed by X-ray broadband ULX spectra obtained with \xmm\ and \nus\ (e.g. \citealt{bachetti13,walton14,walton20}). Moreover, above the critical rate $\dot{M}_{\rm c} = 9/4~\dot{M}_{\rm Edd}$ (\citealt{poutanen07}), powerful outflows of ionised plasmas are expected to be driven by radiation pressure, as shown by magneto-hydrodynamical simulations \citep[e.g.][]{ohsuga11}. These winds imprint weak lines in the high-resolution X-ray spectra of ULXs \citep[e.g.][for a review see also \citealt{pintokosec23}]{pinto16, kosec21}, primarily in the soft X-ray band (0.3--2 keV). 
The discovery of winds in two pulsating ULXs, NGC 300 ULX-1 \citep{kosec18b} and the Galactic Swift J0243.6+6124 \citep{vdEijnden2019}, implies that the dipole component of the magnetic field is $\lesssim10^{13}$ G, otherwise the latter would truncate the disc beyond the spherisation radius ($R_{\rm sph}= 27/4 \ \dot{M} \ R_{G}$) and prevent radiation pressure from launching the observed, relativistic ($0.22$c), wind. Although inclination effects, ionization of the accretion flow or beaming can be not negligible, the discovery of a wind with a certain escape velocity might therefore place constraints on the field strength (for the dipolar component) and the nature of accretor.

\begin{figure*}
\center
\includegraphics[width=17.5cm]{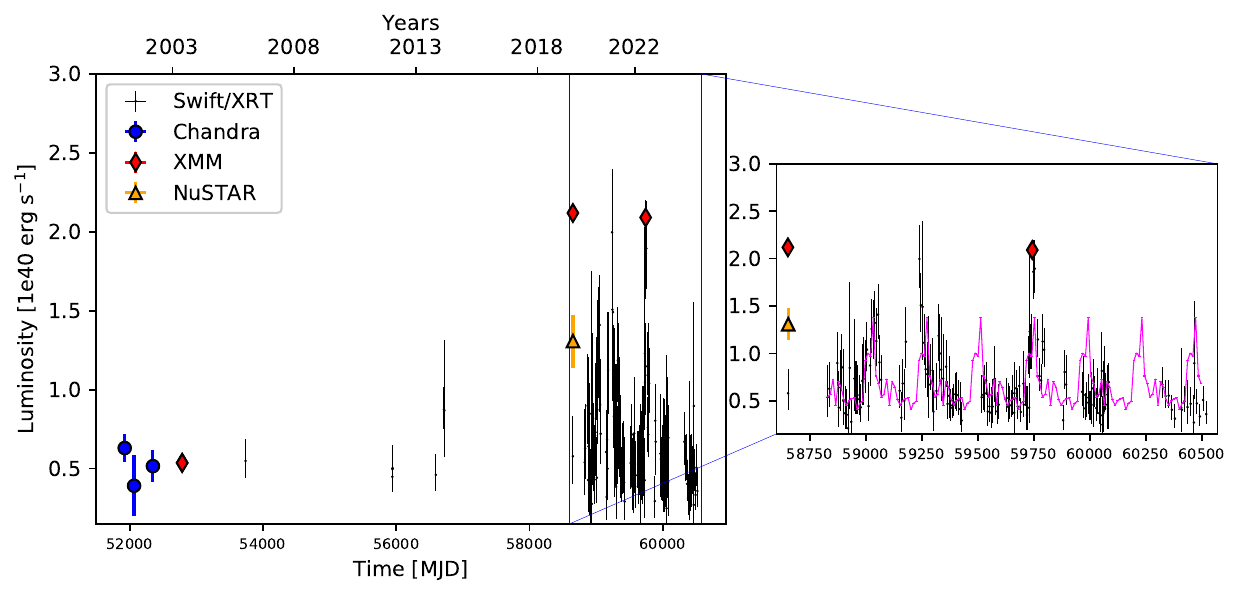}
   \caption{Lightcurve of X7 in the 0.3--10 keV energy band, using all available X-ray observations. The \xmm, \nus, \chandra\ and \swift\ luminosities are not corrected for absorption and are taken as the average in each observation (assuming a distance of 7.5 Mpc). \chandra\ data are taken from \citet{pintore21}. The inset shows the lightcurve after 2018 with the super-position of the profile of the tentative $\sim240$~d super-orbital periodicity (magenta). }
   \label{lcall}
\end{figure*}

The ULX NGC 4559 X7 (hereafter X7), in the galaxy NGC 4559 (distance 7.5 Mpc; \citealt{tully16}), is a peculiar object with a marked flaring activity during the epochs of its highest observed flux states \citep{pintore21}. Such flares were shown to be driven by flux changes at energies higher than 1~keV, with a clear harder-when-brighter behaviour \citep{pintore21}, which is seen in several ULXs \citep[e.g.][]{gurpide21}. High resolution spectroscopy data from \xmm/Reflection Grating Spectrometer (RGS) data revealed a number of narrow absorption and emission lines that are compatible with the existence of an outflow in X7. In addition, variations in the hard band ($2-10$ keV) were seen to lag behind the soft band ($0.3-2$ keV) with a delay of hundreds of seconds \citep[e.g.][]{demarco13, pinto17,kara20}, suggesting that we may be seeing accretion rate fluctuations propagating inwards through the accretion flow. However, the nature of the compact object powering X7 still remains puzzling.

In this work, we extend the analysis shown in \citet{pintore21} and we further investigate the short and long-term spectral and temporal properties of X7, aiming to identify if the system hides a super-Eddington accreting BH or a NS.

\section{Data reduction}
\label{data_reduction}
\vspace{0.1cm}
\subsection{XMM-Newton} \xmm\ observed NGC 4559 on 27 May 2003 for $\sim$42 ks (hereafter XMM0; ObsID: 0152170501, PI: M. Cropper). Then our group obtained a new $\sim$75 ks-long observation (hereafter XMM1; ObsID: 0842340201, PI: F. Pintore), which was taken on 16 June 2019, and a third one on 13 June 2022 of $\sim$130~ks (hereafter XMM2; ObsID: 0883960201, PI: C. Pinto). XMM0 and XMM1 were already reported in \citet{pintore21}. XMM1 and XMM2 caught the source during its highest fluxes (see next sections).

We reduced the observations with {\sc SAS} v20.0.0, using data from both EPIC-pn and EPIC-MOS (1 and 2) cameras: events were selected with {\sc pattern}$\leq$4 (i.e. single and double-pixel events) for the pn, and {\sc pattern}$\leq$12 (i.e. single- and quadruple-pixel events) for the MOS.  All observations are affected by high background time intervals, which were removed from the analysis. This resulted in a net exposure time of $\sim 35$~ks, $\sim 65$~ks and $\sim 95$~ks for XMM0, XMM1 and XMM2, respectively. We extracted source and background events from circular regions with radii of 30'' and 60'', respectively. The photon times of arrival (ToAs) were converted to the barycenter of the Solar System with the task \textsc{BARYCEN}, using the best \chandra\ coordinates of the target (RA=12h 35m 51.71s, Dec=+27d 56m 04.1s; \citealt{swartz11}) and ephemeris DE405{\footnote{https://ssd.jpl.nasa.gov/planets/eph\_export.html}}. 

EPIC spectra were grouped with the FTOOLS {\sc grppha} in order to accumulate at least 25 counts per energy bin. The {\sc rmfgen} and {\sc arfgen} tools were used to generate response matrices and ancillary files, respectively. For all the EPIC lightcurves, we applied the SAS task {\sc epiclccorr} to take into account vignetting, point-spread-function, quantum efficiency and bad pixels. 

\smallskip
{High-resolution spectra require long exposures to estimate the background accurately and to fill the several thousand energy channels. We focused on the data from the RGS of both the XMM1 and XMM2 observations only, because the short exposure of XMM0 was not enough for RGS analysis of its data. 
The RGS data were reduced with the \textsc{rgsproc} task, which produces calibrated event files, spectra and response matrices.}
Following the standard procedures, we filtered the RGS data for solar flares using the background light curve from the RGS CCD number 9 (corresponding to $\lesssim7.5$\,{\AA} or $\gtrsim1.7$\,keV). The background light curves were binned at a 100~s time resolution and all the time bins with a count rate above 0.2 cts s$^{-1}$ were rejected, \noindent {using the same good time intervals (GTI) for the RGS 1 and 2 detectors. We extracted both the $1^{\rm st}$- and $2^{\rm nd}$-order RGS spectra in a cross-dispersion region of 1 arcmin width, centred on the same source coordinates used for extracting the EPIC spectra. The background spectra were extracted from regions outside the 98\% of the source point-spread-function. Due to the possible presence of further fainter sources in the background regions, we also extracted the background spectrum produced using the template background model from blank field data
(obtained from the {\textit{rgsproc}} pipeline with the flag `withbackgroundmodel=yes'). The two background spectra were consistent. 
In order to obtain the highest signal-to-noise ratio, we stack the $1^{\rm st}$- and $2^{\rm nd}$-order spectra from the two observations for a total of 176 ks for each spectral order. Such a combined RGS spectrum provides about 9400 net counts, which broadly agrees with the $\gtrsim 10,000$ counts typically needed to significantly detect lines in ULXs (\citealt{pinto16,kosec21}). To search for spectral features, we also need an accurate knowledge of the continuum shape, especially for photoionisation modelling. We therefore stacked the MOS\,1, MOS\,2, and pn spectra of XMM1 and XMM2 to cover the 2-10 keV band. In the end, we have 5 combined spectra for the XMM1 and XMM2 observations (RGS $1^{\rm st}$- and $2^{\rm nd}$-order, MOS1, MOS2 and pn). 

\subsection{Swift} The XRT instrument onboard the {\it Neil Gehrels Swift Observatory} (hereafter \textit{Swift/XRT}) observed the galaxy NGC 4559 for 164 times between January 2006 and July 2024 (Table~\ref{list_of_log}). We note that the vast majority of the observations were taken in Photon Counting (PC) mode with an average exposure time of about 2 ks each, generally taken with a weekly or bi-weekly cadence. We excluded all the observations in other modes (e.g. windowed timing) in order to avoid contamination from other sources in the field of view.

\noindent We reduced all observations with the {\sc xrtpipeline} tool of the {\sc ftools} software, and we extracted source and background events from circular regions of radii 40'' and 60'' (centered at the source position and in a source-free region of the CCD), respectively. The long-term lightcurve of the \swift\ observations is shown in Fig.~\ref{lcall}. For a comparison with the \xmm, \nus\ and \chandra\ data, we converted \swift\ count rates in luminosities. Due to the low statistics of the single \swift\ observations, we performed the count-rate to flux conversion by firstly fitting the stacked spectrum\footnote{The stacked spectrum was generated through the online tool \url{https://www.swift.ac.uk/user_objects/} of \citet{evans09}} of all \swift\ observations with an absorbed power-law spectral models. We determined the best-fit by fixing the column density to $1.5\times10^{21}$ cm$^{-2}$, which is the value found from the high quality XMM spectra (see Section~\ref{spec_analysis}), and let the power law photon index $\Gamma$ and normalization free to vary. The best fit of the stacked spectrum gives $\Gamma\sim2.2$ and 0.3--10 keV flux of $1.1\times10^{12}$ erg cm$^{-2}$ s$^{-1}$. We adopted this model, freezing the photon index and letting free to vary only the normalisation, to estimate the observed 0.3--10 keV source luminosity in each \swift\ observation.

\section{Results}

\subsection{Long-term analysis}
\label{long_term}

As reported in \citet{pintore21}, a tentative super-orbital periodicity of $\sim200$~d was proposed to describe the (still limited at that time) \swift\ long-term lightcurve of X7. Thanks to the regular \swift\ monitoring performed between 2020 and 2024, the baseline to search for periodicities is significantly increased.

\begin{figure}
\center
\includegraphics[width=9.4cm]{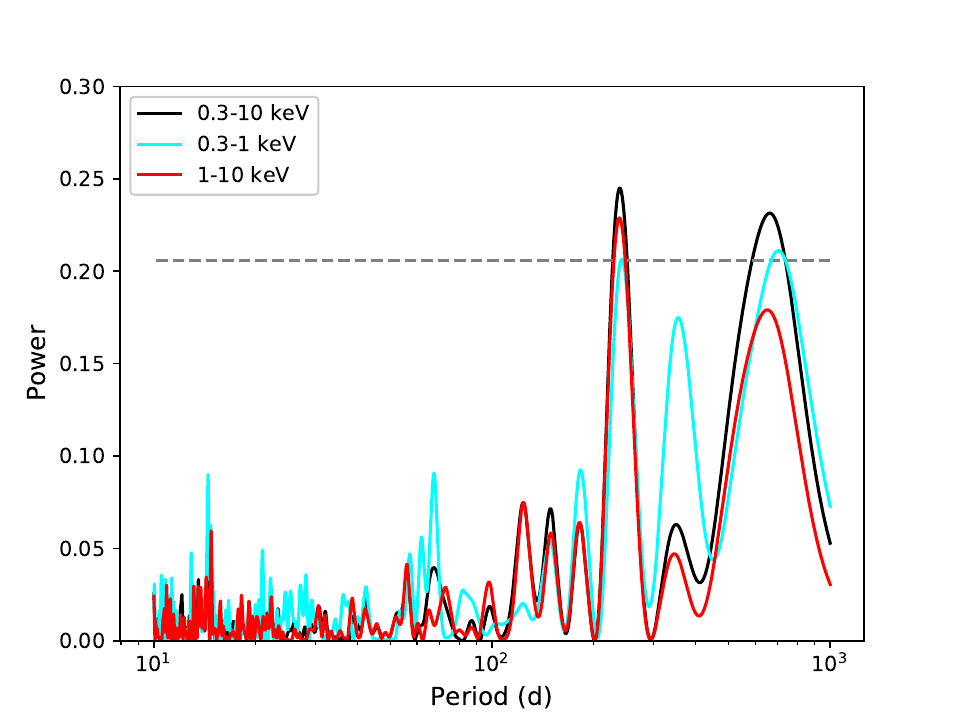}
\hspace{-1.9cm}
\includegraphics[angle=270, width=9.cm]{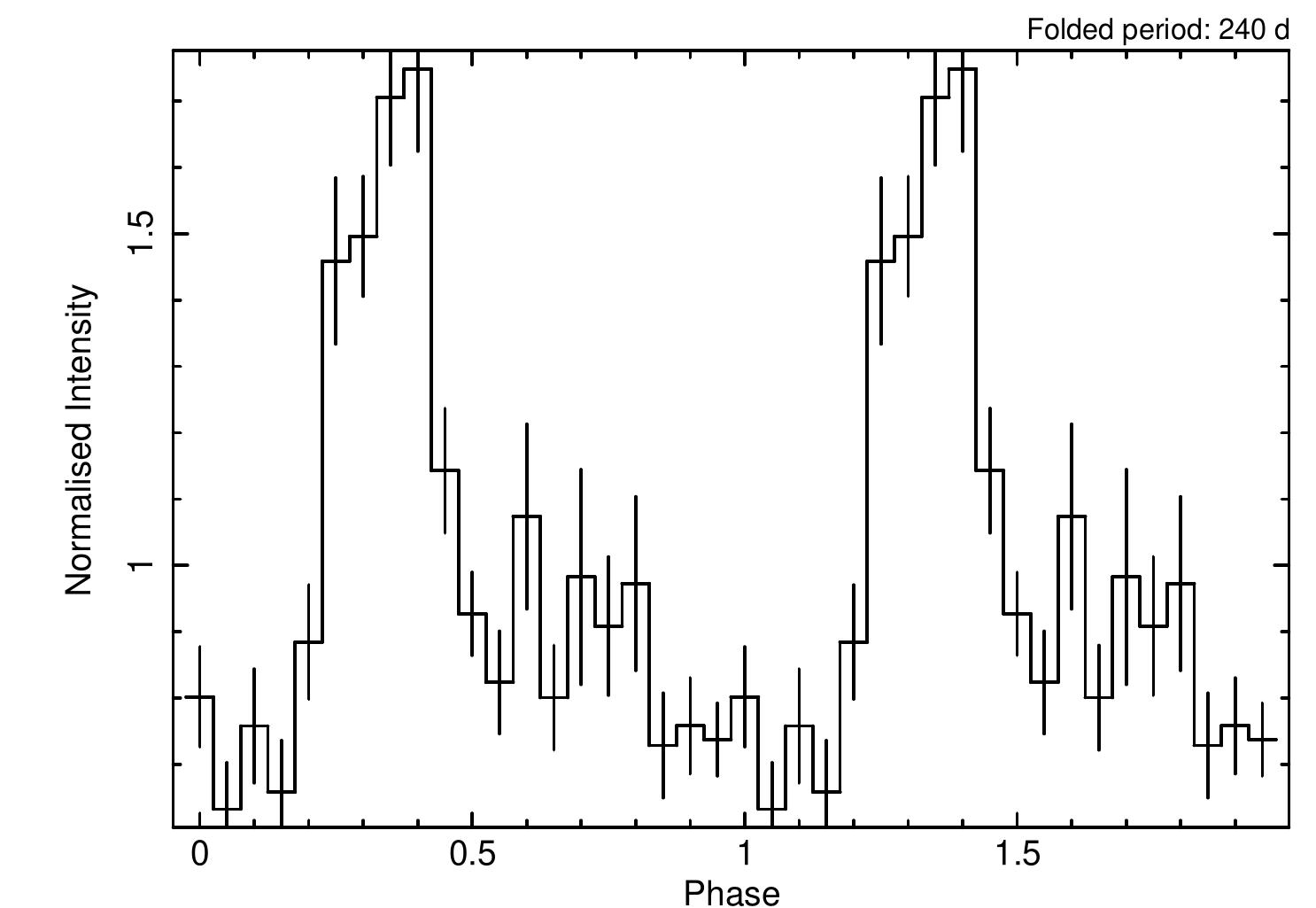}
\caption{{\it Top}: Lomb-Scargle periodograms of the \swift\ observations in the time intervals between 2020 and January 2023 in the 0.3--10 keV, 0.3--1 keV and 1--10 keV energy range. Dashed lines indicate signals with $>3\sigma$ significance. {\it Bottom}: The full 1--10 keV \swift\ lightcurve folded with a period of $240$~d.}
\label{LS}
\end{figure}

\begin{figure*}
\center
\includegraphics[width=18.3cm]{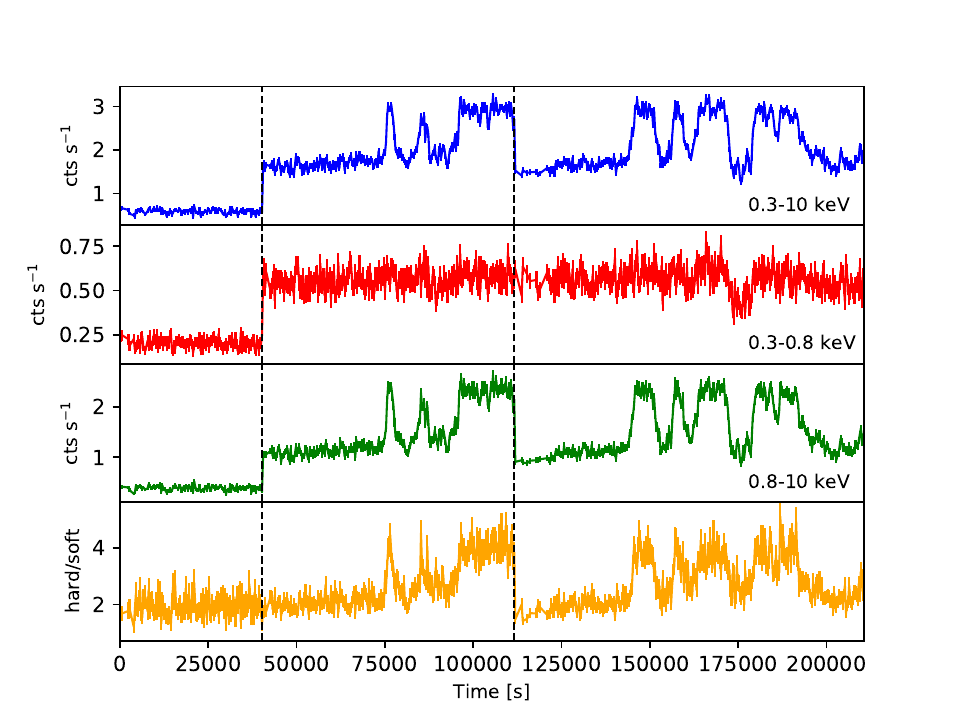}
   \caption{Background subtracted lightcurve of X7 in XMM0, XMM1 and XMM2 (from left to right), binned at $\Delta T = 300$s (see the top-left legend), for the 0.3--10 keV ({\it top panel}), 0.3--0.8 keV ({\it center-top}),  0.8--10 keV ({\it center-bottom}) energy bands. We also show the hardness ratio between the hard and the soft band ({\it bottom panel}). The dashed, vertical lines divide the data from XMM0, XMM1 and XMM2, where the real time gaps between them were removed for displaying purposes only.}
   \label{lightcurve}
\end{figure*}

We ran a Lomb-Scargle (LS) analysis over all observations between 2020-2024 and investigated the energy bands 0.3--10 keV, 0.3--1.0 keV and 1.0--10 keV, with a bin time of 2 days. We used the {\tt LombScargle} class of the python {\it Astropy} library, where we set the method to initially estimate the false alarm probability\footnote{The probability of measuring a peak of a given height (or higher), conditioned on the assumption that the data consists of Gaussian noise with no periodic component.} of any signal to {\it Baluev} \citep{baluev08}.
The LS, applied separately to the three energy bands, revealed an indication for a periodicity in the range 500-600 d with a 3$\sigma$ significance in the total and the 0.3--1.0 keV bands.
We note that the peak is very broad, pointing towards large uncertainties. In addition, such a periodicity does not cover more than 2-3 cycles of the whole lightcurve, suggesting that it should be only treated as a possible spurious signal due to the red noise.} 

On the other hand, we note that excluding the observations after 21 January 2023 (MJD 59965), i.e. during which the source showed a prolonged and stable low flux, the LS provides again the periodicity of $\sim500-600$~d but also signals at 239~d (significance $>3\sigma$), 238~d (significance $>3\sigma$), and 242~d (slightly less than $3\sigma$) in the 0.3--10 keV, 1--10 keV and 0.3--1.0 keV energy bands, respectively (Fig.~\ref{LS}-top). 

We assess the uncertainty on the $\sim240$~d periodicity by performing a Monte Carlo simulation. For each point of the real \swift\ lightcurve, we sampled a random value following a normal distribution with mean and standard deviation equal to the observed X7 count-rate and its error, respectively. With this approach, for a given energy band, we created a simulated lightcurve with the same length as the original one, we performed a Lomb-Scargle analysis on it and we determined the main peaks of the periodogram. We simulated 10000 lightcurves in the three energy bands and drew the distribution of the Lomb-Scargle peaks. For each peak, we estimated the mean and the standard deviation of its distribution by fitting it with a Gaussian model. The standard deviation was finally associated to the uncertainty on the periodicities determined from the observed lightcurves. The uncertainties (1$\sigma$) on the periodicity for the 0.3--10 keV, 1--10 keV and 0.3--1 keV are 2.4~d, 2.4~d and 3.4~d, respectively, which makes all the periodicities widely compatible with 240~d within less than 2$\sigma$. We used 240~d as reference time for the next analyses.  

{ Should the shorter periodicity at $\sim240$~d be real} in these data, we note that it is unlikely to be sinusoidal in shape. The agreement of the best-fit sinusoidal curve and the X7 lightcurve is not very suitable for the high flux states of the source and it tends to deviate at the epochs where the Lomb-Scargle fit was not performed. To confirm the non-sinusoidal behaviour, we run also an epoch folding search with the {\sc FTOOLS} task {\sc efsearch} in the time range before January 2023 of the 1--10 keV lightcurve. We searched around a period of 240~d, binning the curve with 20 bins and with a time resolution of 10000~s. Although the lightcurve has numerous time gaps which can introduce spurious signals, we found the strongest periodicity at 240.15~d. Fitting the $\chi^2$ profile with a constant plus a gaussian, we estimated that the error on the period is 0.1~d at $1\sigma$. This value is therefore compatible with the LS one. We folded all the \swift\ observations in the 0.3--10 keV band with the a 240~d periodicity (taken, as average value; see Fig.~\ref{LS}-bottom), indicating that the folded profile presents a peaked signal followed by a flux decay. In Fig.~\ref{lcall}, we show the complete X7 long-term lightcurve where, in the inset, we overlaid the profile of the $\sim240$~d periodicity.

In order to test the statistical significance of the signal found with the LS, we firstly performed an approach based on scrambling the data of the lightcurve. Specifically, we assigned each observation a count rate taken from another, random observation; on the new generated curve, we performed a LS analysis and we took the maximum peak power in the period range of our signal. We created 10000 new lightcurves for the 1--10 keV energy range (taken as reference) and we found that the number of signals with a peak power equal or higher than 0.229 (the power found in the real data for the periodicity) was $<10$, implying that the $\sim$240~d observed signal is significant at $>3\sigma$.

However, we note that the scrambling of the date and the false alarm probability of the LS do not take into account any red-noise, that can be important with long-term periodicities \citep{scargle81} and can lower the significance of the signals \citep{vaughan03}. We further created a set of 10000 simulations following the approach presented in Section 3 of \citet{walton16}. We simulated a light curve assuming that the power spectrum has a power law shape with index of 1 or 2, with the index values being a guess on the real slope.  
The light curves were simulated with a time bin of 2 ks, i.e. the average exposure of the \swift\ monitoring. The length of the simulated light curve was increased to 20 times the observed light curve. Then, we picked only the time bins consistent with the times of the real \swift\ observations. For each filtered lightcurve, we performed a LS analysis. We then determined how many peaks have power equal to or larger than the one found in the observed data. We calculated it for the 1--10 keV energy band, as reference. We found that, between 200~d and 300~d, a total of 87 and 1042 peaks over 10000 were found above a power of 0.229 for a power law index equal to 1 and 2, respectively. Hence, in the presence of red noise, our results would imply that the actual significance of the 240~d periodicity could be lower than the $3\sigma$ reported above.

These tests prove that the real nature of the periodicity cannot be verified yet and more observations are necessary.

\subsection{Short-term temporal analysis}
The XMM1 and XMM2 observations of X7 show that, in both epochs, the source was in a similar regime, with strong and repeated flares (Fig.~\ref{lightcurve}). The flares enhance the observed emission of a factor of 2 with respect to the persistent level and, more interestingly, they all appear flat-topped at a very similar flux. We note that the duration of the flares is not regular and varies from a few ks to tens of ks.
\newline
\newline
{\bf Time lags. } We performed a search on time lags for XMM2, over the same timescales and energies as those reported for XMM1 in \citet{pintore21}. However, no significant lags were found, implying that the hard lag detected in XMM1 is not observed again. 
\newline
\newline
{\bf Search for quasi periodic oscillations. }{We searched for quasi-periodic signals (QPO) in XMM2\footnote{For XMM1, the search for QPOs was reported in \citet{pintore21}.} data by creating power density spectra (PDS) from segments of 3000~s each, resulting in a frequency resolution of approximately 0.0003 Hz. We selected the length of these intervals to ensure sufficient photon accumulation for an adequate signal-to-noise ratio, while also preserving any potential transient signals. We produced a total of 27 PDS.

In some PDS, a narrow feature at about 1~mHz was observed, although it was not statistically significant. Similar features are sometimes visible in photon-starved EPIC-PN datasets, which may indicate that they are either statistical fluctuations or instrumental artifacts.
We also combined the 27 PDS into a single averaged PDS to reduce uncertainties. {This choice is supported by the fact that each PDS is very poor in statistics and there are no significant differences between the PDSs extracted in segments containing either the flares or the persistent emission. The PDSs are all dominated by noise, implying no significant variability over timescales $\leq 3$ ks.} The averaged spectrum is dominated by Poissonian noise as well, although an unresolved peaked feature is visible at roughly 0.5 mHz. This frequency is very close to our frequency resolution, so we conclude that it is unlikely to be a real signal.}
\newline
\newline
{\bf Pulsation search.} We then conducted a search for coherent signals from X7 in XMM2\footnote{We did consider the full observation, i.e. not excluding high background epochs) as they weakly affect the search for pulsations}. 
\begin{figure}
\center
\includegraphics[width=9cm]{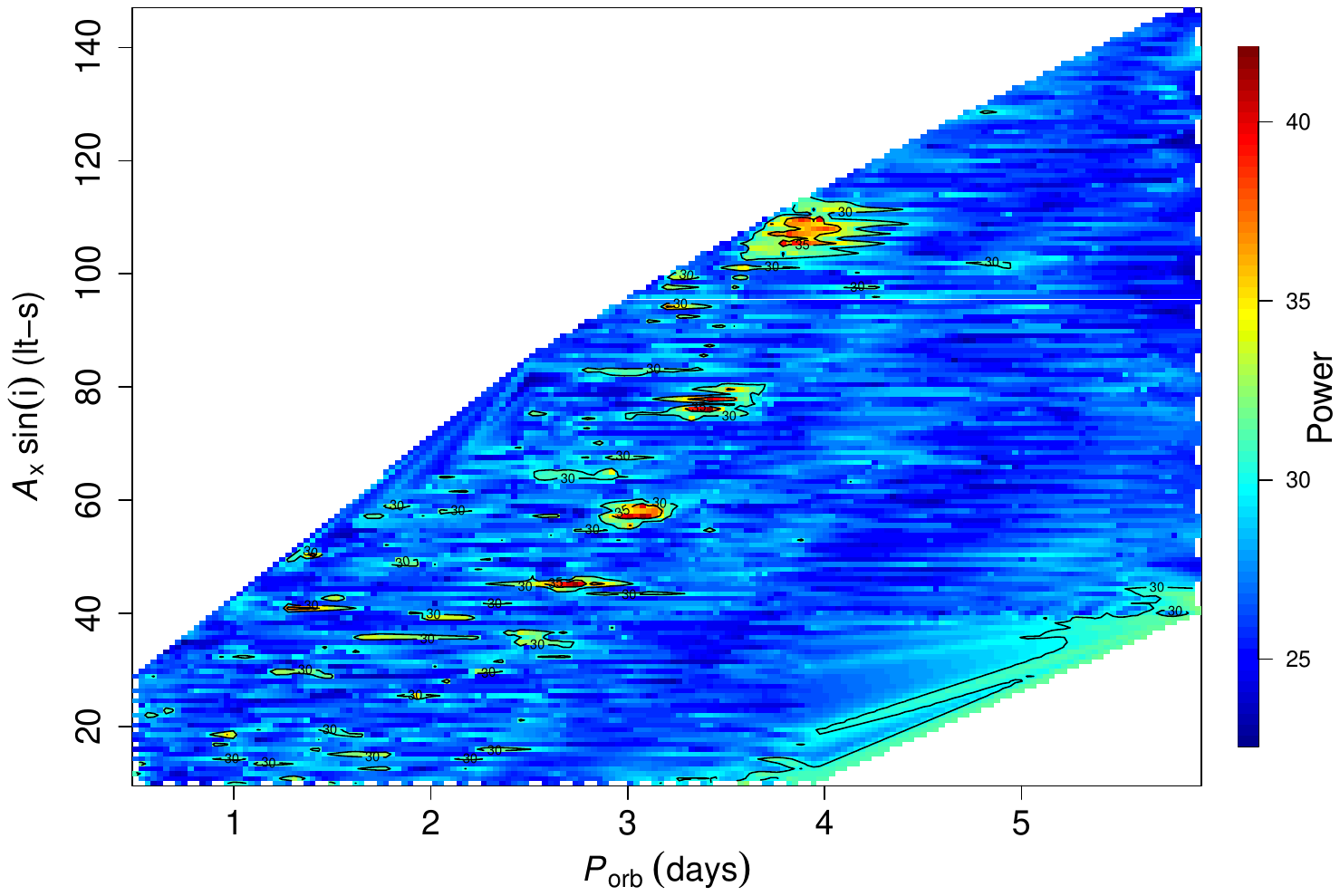}
   \caption{Estimates of the X7 orbital period ($\text{P}_{\text{orb}}$) and the projected semi-axis of the orbit for a spin period of $\sim2.73$s in XMM2. The confidence contour, more similar to an atoll rather than a ``banana'' shape, is indicative of the low significance of the spin period detection. However, a similar plot can be obtained for XMM1 as well, boosting the real significance of the signal. { The colourbar indicates the Leahy power of the signal.}}
   \label{sopa}
\end{figure}
We performed accelerated searches for periodic signals: to account for potential variations in spin period due to intrinsic spin-down or spin-up, the proper motion of the system, or the orbital motion of the compact object, we initially corrected the arrival times of the events using a grid of approximately 930 $\dot{P}/P$ values within the range of ($\pm$) 10$^{-11}$ s s$^{-2}$ to 10$^{-5}$ s s$^{-2}$. We note that this does not take into account any orbital correction. The search did not yield any statistically significant signals and resulted in a $3\sigma$ upper limit of 5\% on the fractional amplitude of a signal within the period range of 0.146 to 100 seconds.

Subsequently, we conducted an extensive search by applying binary-system orbit corrections to the photons ToA, accounting for the motion of the compact object around its companion. This involved the analysis of $\sim30,000$ different orbital configurations with orbital periods ranging from 4 hours to 4 days and orbital separations ranging from 2 to 120 light-seconds. 
A marginally significant signal was detected over several orbital corrections corresponding to a spin period $P\simeq$ 2.73 s (Fig.~\ref{sopa}). Its statistical significance could be reduced further due to the large number of trials. However, the large number of orbital parameters that gave rise to this signal detection encouraged us to search for such a signal in the XMM1 data. In fact, we did detect a quite close signal at $P\simeq$ 2.63~s in this observation. Even though the detected signal in XMM1 was also marginally significant, the result that a very similar signal was detected in two separate observations improves its statistical significance. Taking into account that the signal has a very low pulsed fraction\footnote{defined as (F$_{max}$ - F$_{min}$)/(F$_{max}$ + F$_{min}$), where F$_{max}$ and F$_{min}$ are the maximum and minimum values of the observed photon flux} (PF) of $\sim$5\%, it means that a low power in the PSD is expected. The orbital parameters are quite unconstrained, providing an orbital period in the range 2--4~d and $A_{x}~\text{sin}(i)$ of $\sim$20--120 lt-s.
The detected signals would imply a quite strong, three-year spin-down rate of $\dot{P}/P \sim - 1.06 \times 10^{-9}$ s s$^{-1}$. Still, the maximum significance of the signals is close to $\sim 3.5 \sigma$, which does not allow us to firmly claim it is a real signal. Therefore, we present it as a possible/candidate spin period. Further observations are needed to confirm its significance.
\newline
\newline
{\bf Pulsation search with evolutionary algorithms.} We also conducted another search for coherent signal in XMM1 and XMM2 using \texttt{stingray} (version 2.0.0) and Particle Swarm Optimization (PSO), an evolutionary algorithm \citep{Kennedy}, similar to the search presented in \citep{Sacchi2024}. PSO is a meta-heuristic algorithm designed to find the maximum value of a dependent variable in a multivariate function. In our case, the independent variables are the two orbital parameters ($A_{x}~\text{sin}(i)$ and orbital period), $\dot{P}/P$, and the phase, while the dependent variable is the power of the signal. These independent variables define a $\mathbb R^4$ space (the search space). More details about the method are described in Appendix A.

\begin{figure*}
\center
\includegraphics[width=8.8cm]{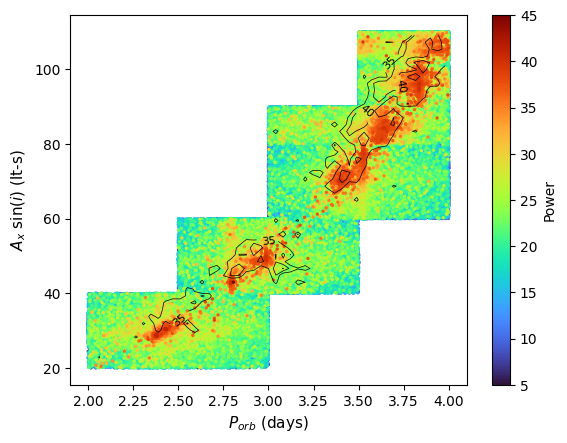}
\includegraphics[width=8.8cm]{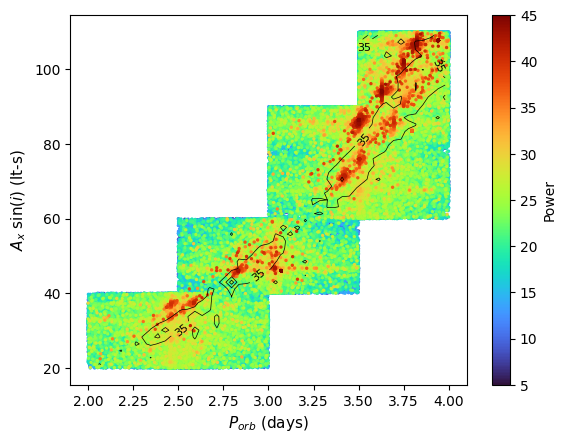}
\caption{Parameter space of the orbital parameters for the signal at 2.63s and 2.73s for XMM1 (left) and XMM2 (right), respectively. For displaying purposes only, in each plot the contours indicate the solution of the other observation. {The colourbar indicates the Leahy power of the signal.}}
\label{fig:pso_power}
\end{figure*}

The parameter space for $A_{x}~\text{sin}(i)$ and the orbital period P$_{\text{orb}}$ present an evident ``banana'' shape for two signals at $\sim2.63$s and $\sim2.73$s for XMM1 and XMM2, respectively, confirming hence the results obtained with more standard algorithms. These signals are associated to a maximum power of 47.99 for XMM1 and 45.27 for XMM2. We observed that high power values show a wide range of $\dot P/P$ ([$-1 \times 10^{-12}, 2 \times 10^{-11}$] s s$^{-1}$ and [$-3 \times 10^{-12}, 1 \times 10^{-10}$] s s$^{-1}$ for XMM1 and XMM2, respectively), which implies that $\dot P$ cannot be constrained. On the other hand, we found that the uncertainties (at $1\sigma$) for the coherent signals are $P_{spin} = 2.628\text{s} \pm 0.001$s for XMM1 and $P_{spin} = 2.726\text{s} \pm 0.012$s for XMM2. 

It is important to note that the shapes of the two ``bananas'' obtained by these signals are totally overlapping (Fig.~\ref{fig:pso_power}). 
Such a result suggests that the two signals may have a common origin since, if they were fake, it would be extremely unlikely to observe the same parameter space for the orbital parameters. To investigate if the coherent signal can be an artefact, we selected the largest number of background events ($\sim30000$) from XMM1 taken from the same CCD of the source and we performed again the analysis on these events. We did not find any signal at the same frequencies with powers comparable with (or higher than) those found for X7. The results support and strengthen the presence of pulsation in the source.

\subsection{Hardness-intensity analysis}
\label{hid_sec}

\begin{figure}
\center
\includegraphics[width=9.cm]{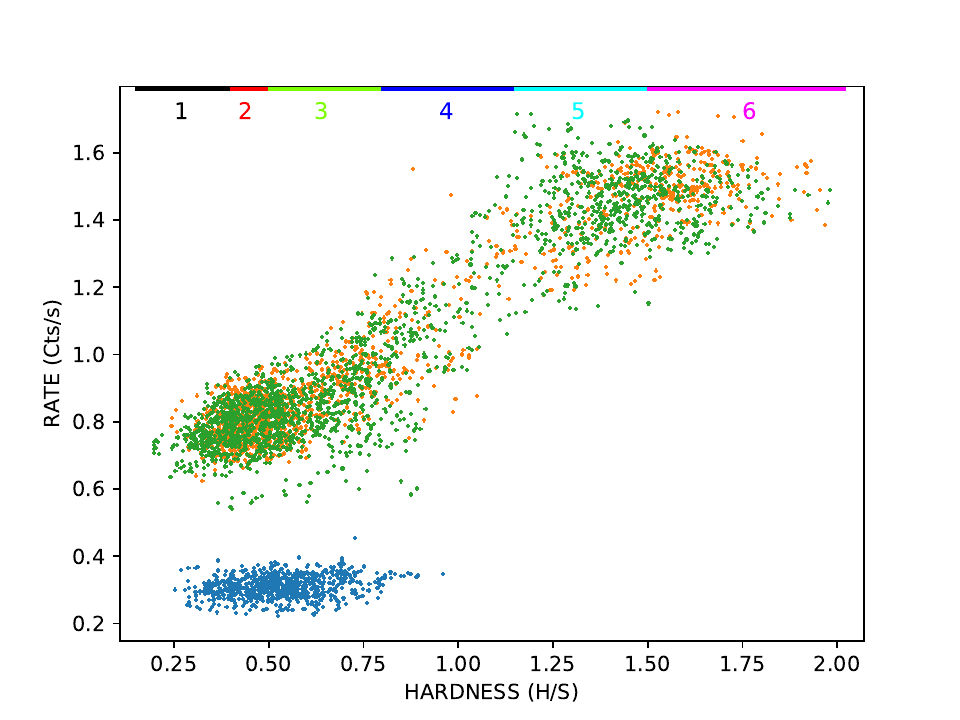}
\includegraphics[width=9.cm]{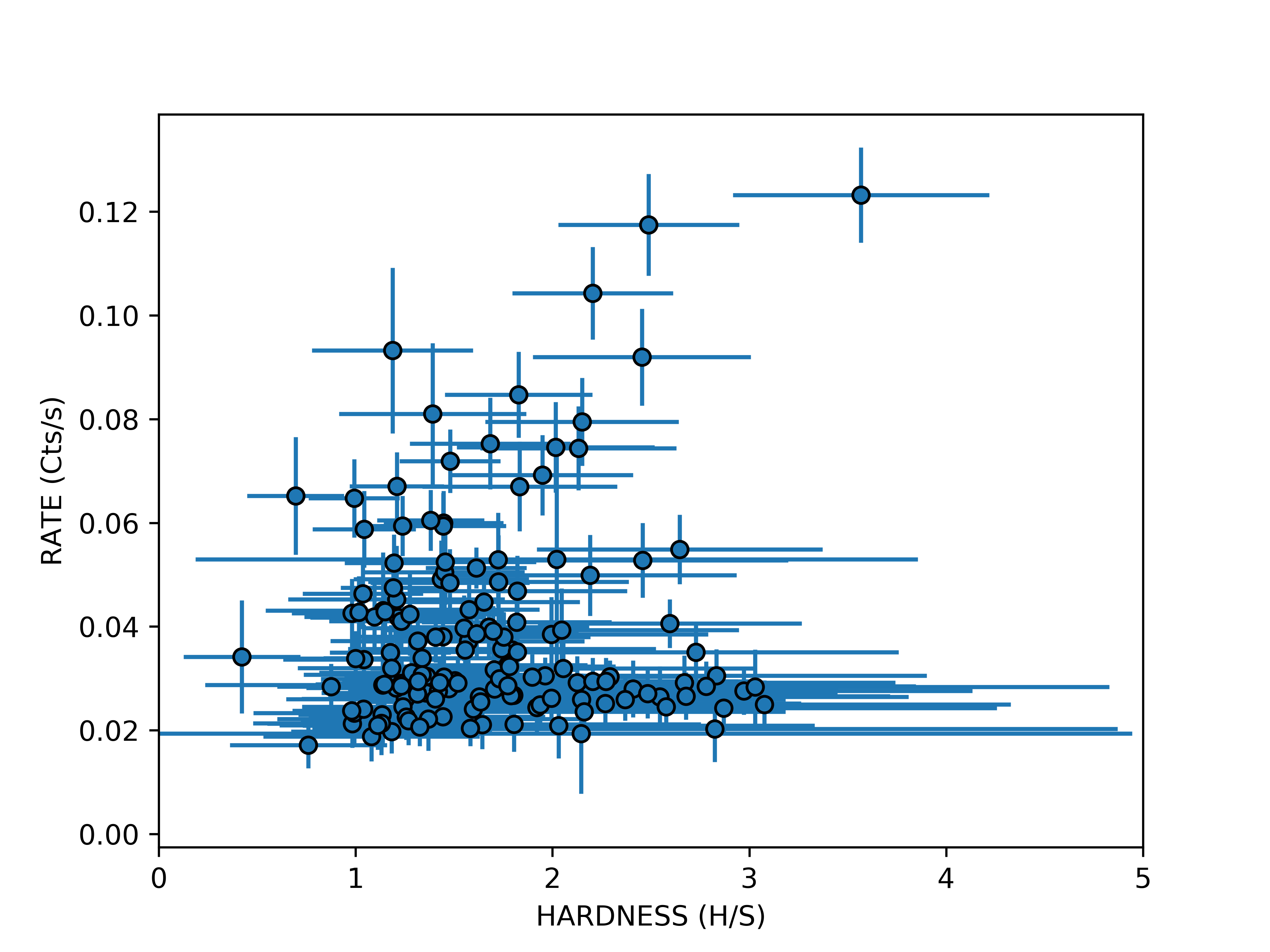}
   \caption{{\it Top}: Hardness-intensity diagram for PN+MOS lightcurves of all \xmm\ observations (after applying a smoothing to the data, see text). The data of the XMM0, XMM1 and XMM2 observations are reported in blue, orange and green, respectively. Hardness is defined as the ratio between the bands 2.0--10 keV (H) and 0.3--0.8 keV (S), while the rate is the sum of both bands. Smoothing with a moving window has been applied to the data (see text) and therefore we do not plot the errors on each point. The coloured and numbered segments on the top indicate the hardness ranges used for the spectra extraction shown in Fig.~\ref{spec}.
   {\it Bottom}: HID for the available \swift\ data; hardness is the ratio between the bands 1.0--10 (H) keV and 0.3--1.0 (S) keV. 
   }
   \label{hid}
\end{figure}

We built a hardness-intensity diagram (HID) with all the \xmm\ observations by using the EPIC data (pn + MOS1, 2) in the energy bands 0.3--0.8 keV and 2.0--10 keV (Fig.~\ref{hid}-top). The two bands are slightly different than those used in the previous section: { with this choice, we can evaluate the HID using energies where we are confident the low and high energy spectral spectral components of X7 dominate { (i.e. below 1 keV and above 2 keV; \citealt{pintore21})} and, therefore, to maximize the possibility to find evidence of well distinguished spectral states.}
The background subtracted lightcurves, binned with $\Delta T=50$s, were smoothed with a moving window method in order to reduce the noise fluctuations of the data, following the same approach proposed in \citet{dai22}. The HID shows the existence of at least three spectral regimes (Fig~\ref{hid}-\textit{top}), two sharing a similar hardness but different fluxes and one which is both harder and brighter than the others. We labeled the lowest luminosity state (which represents entirely the observation XMM0) as {\it low state}, and {\it persistent state} and {\it flaring state} the other two, respectively. The latters are populated in a very similar way in both XMM1 and XMM2. We also mention the existence of a branch connecting the {\it persistent state} and the {\it flaring state}, which indicates a smooth transition between these two states (albeit a rapid one, given the low density of these data points).

We checked if this behaviour is also observable in the \swift\ data. Because of the low count-rate in \swift, we were forced to use the energy bands 0.3--1.0 keV and 1.0--10 keV to { have enough counting statistics in both bands}. Despite this precaution, the uncertainties due to the poor statistics are still large and do not allow us to robustly constrain the existence of the three regimes (Fig.~\ref{hid}-bottom); however, we note that a similarity between the \swift\ and \xmm\ HIDs can be found, although we warn the reader that the baseline of all the \swift\ observations cover a significantly longer time range than the single \xmm\ observations. Furthermore, we highlight that it might be possible that flares similar to those observed in \xmm\ were caught by \swift\ only partially due to the short exposures (in general, around 2 ks), affecting the final shape of the HID. { Finally, the different energy bands used here and the sensitivity of each instrument can also be responsible for a different HID in \swift.} 

\subsubsection{Spectral analysis}
\label{spec_analysis}
\begin{figure}
\center
\includegraphics[width=9.8cm]{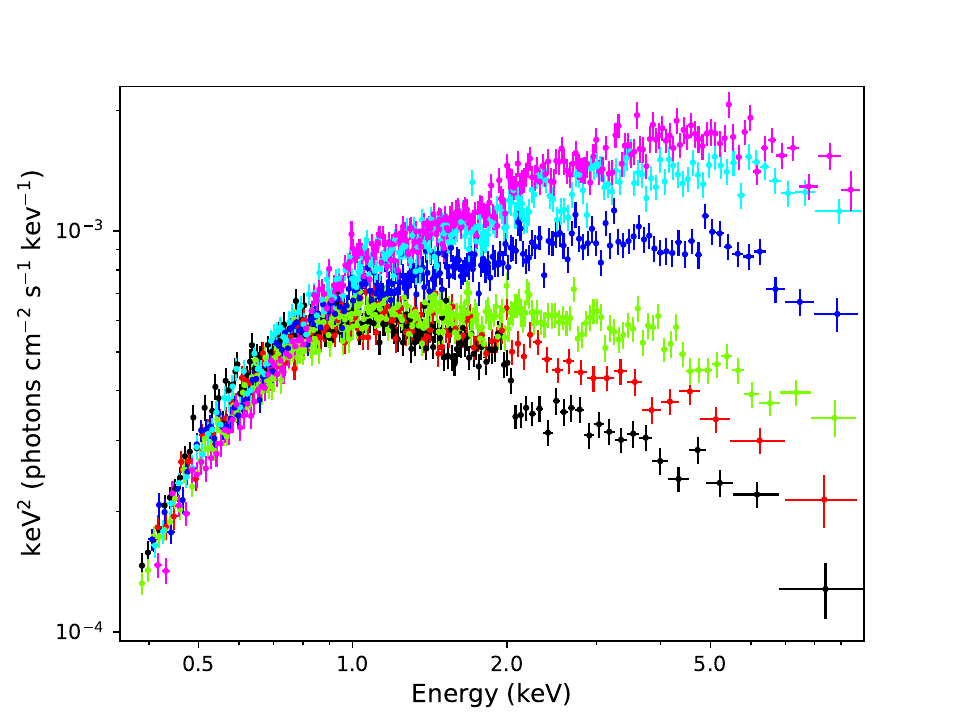}
   \caption{{ Combined EPIC-pn + MOS(1,2) spectra of the six spectral states extracted from the HID (see text), unfolded through a powerlaw of index 0. The colour convention is that indicated in Fig.~\ref{hid}}-{\it top}.}
   \label{spec_all}
\end{figure}

We focus our spectral analysis on the high-quality \xmm\ observations. The HID tracks very well the rapid spectral evolution of the source over time and allows us to investigate in detail the flare peaks, the persistent emission and the transition phase between these two regimes.
Because the X7 spectral properties in XMM0 were already reported in \citet{pintore21} and are not particularly informative, we focussed only on XMM1 and XMM2. We selected six time intervals associated to hardness $<0.4$ (the persistent emission), [0.4--0.5], [0.5--0.8], [0.8--1.15], [1.15--1.5] (the transition) and $>1.5$ (the flares), and we extracted the spectra for both XMM1 and XMM2. This reflects to a total of 12 EPIC-pn and 24 EPIC-MOS (1 and 2) spectra. After checking that similar spectral states have consistent pn and MOS spectra, we combined them with the {\sc sas} task {\sc epicspeccombine}, creating a final number of 6 EPIC (pn + MOS1, 2) spectra (Table~\ref{6states_tab} for details). We performed the fits with {\sc Xspec} v.12.13.0c \citep{arnaud96} in the energy range 0.3--10 keV. { In Fig.~\ref{spec_all}, we plot the six spectra unfolded with a powerlaw of index 0 in order to clearly show the source spectral evolution.}

\begin{table}
\caption{Properties of the stacked EPIC spectra in the six states analysed in this work and corresponding to the numbering of the coloured boxes in Fig.~\ref{hid}.}
\label{6states_tab}
\center
\scalebox{0.9}{\begin{tabular}{lcclc}
\hline
N\# & Net count rate & Tot. net counts & Hardness & Exposure \\
 & cts s$^{-1}$ &  &  & ks \\
\hline
1 & $1.333 \pm 0.007$ & 41171 & $< 0.4$ & 30.4 \\
2 & $1.385 \pm 0.009$ & 24510 & [0.4, 0.5] & 17.5 \\
3 & $1.460 \pm 0.006$ & 58967 & [0.5, 0.8] & 39.9 \\
4 & $1.830 \pm 0.009$ & 42187 & [0.8, 1.15] & 22.9 \\
5 & $2.21 \pm 0.01$ & 44453 & [1.15, 1.5] & 19.9 \\
6 & $2.31 \pm 0.01$ & 50922 & $> 1.5$ & 21.9 \\
\hline
\end{tabular}}
\end{table}

We described the data with a model already reported as suitable for X7 \citep{pintore21}, i.e. the combination of two thermal components and a cut-off power law ({\sc cutoffpl} in {\sc Xspec}), absorbed by a neutral column density ({\sc tbabs} in {\sc Xspec}). The former components are a {\sc diskbb} and {\sc diskPbb}\footnote{The model allows us to fit the radial dependency of the inner disc temperature ($T_{disc}\propto r_{in}^{-p}$), with $p$ assuming the value 0.75 for a standard accretion disc and 0.5 for an advection dominated disc.} model \citep[e.g.][]{mineshige94,kubota04}. The cut-off power law was found in a number of ULXs, and it might hint to an accretion column above a magnetized NS (although other scenarios are possible; e.g. \citealt{mills23}). 
{ We note that the additional cut-off power law was also found to be necessary in the spectra of NGC 4559 X7, when a combined \xmm\ and \nus\ spectral analysis was performed \citep{pintore21}.}

\begin{figure*}
\center
\includegraphics[width=6.4cm]{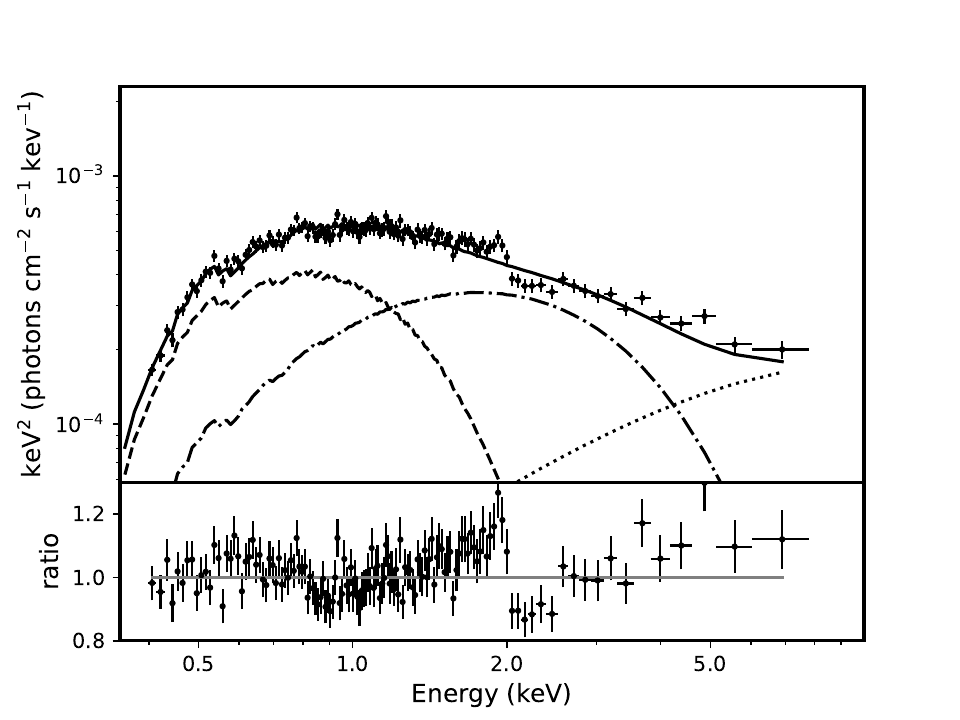}
\hspace{-0.6cm}
\includegraphics[width=6.4cm]{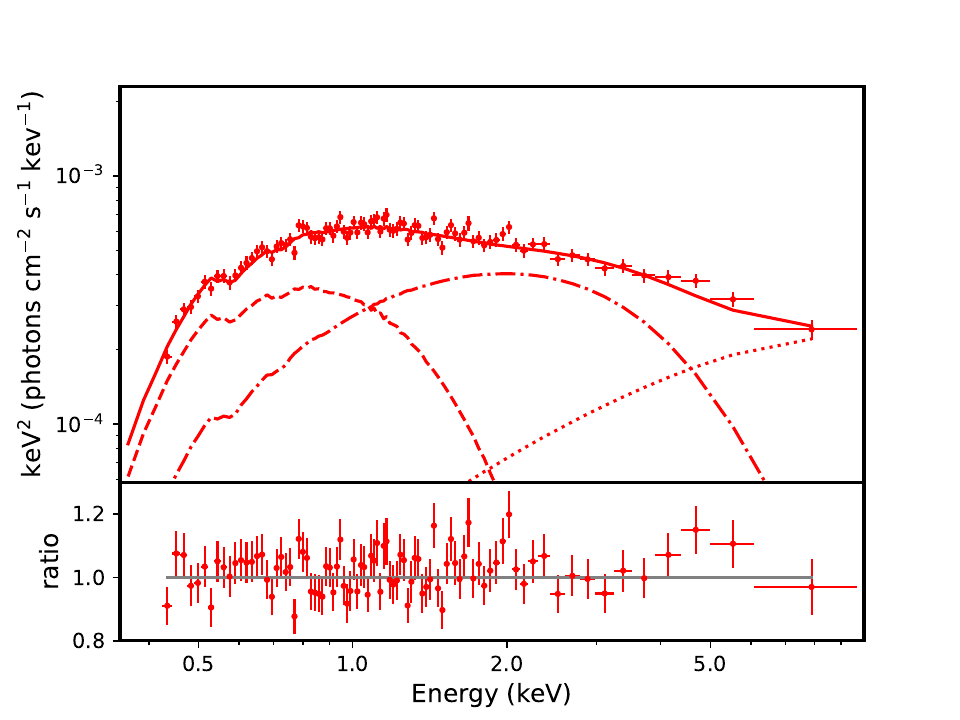}
\hspace{-0.6cm}
\includegraphics[width=6.4cm]{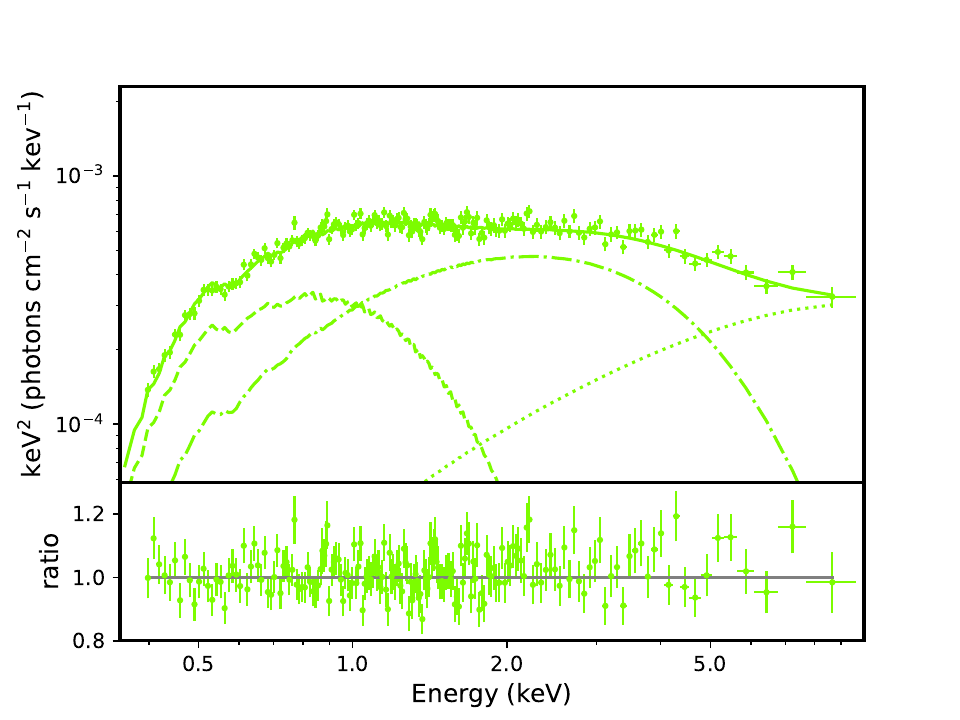}
\hspace{-0.6cm}
\includegraphics[width=6.4cm]{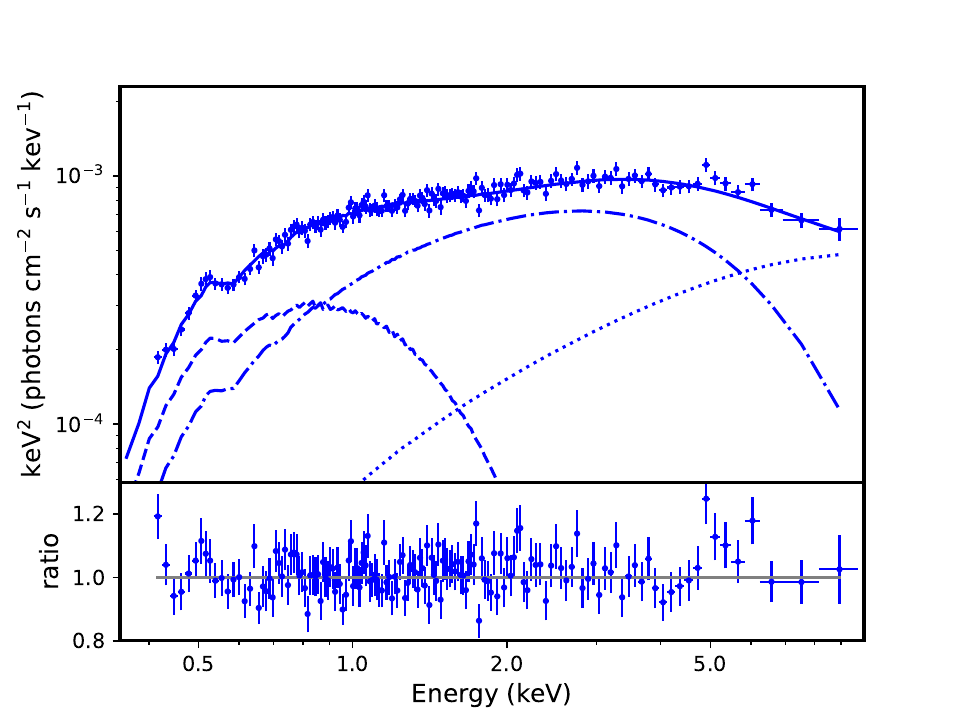}
\hspace{-0.6cm}
\includegraphics[width=6.4cm]{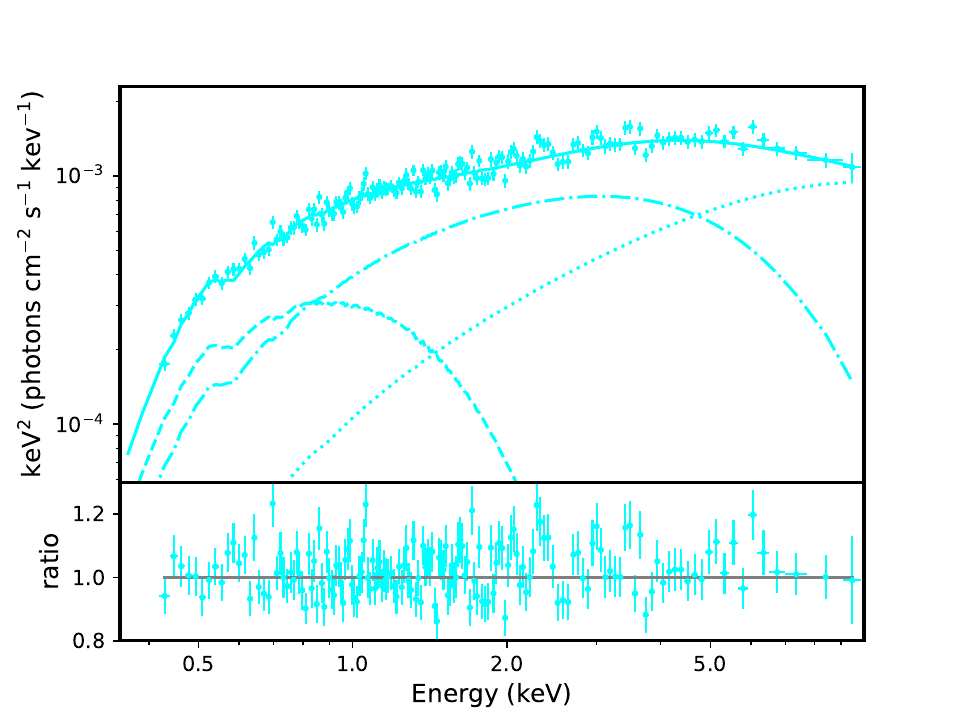}
\hspace{-0.6cm}
\includegraphics[width=6.4cm]{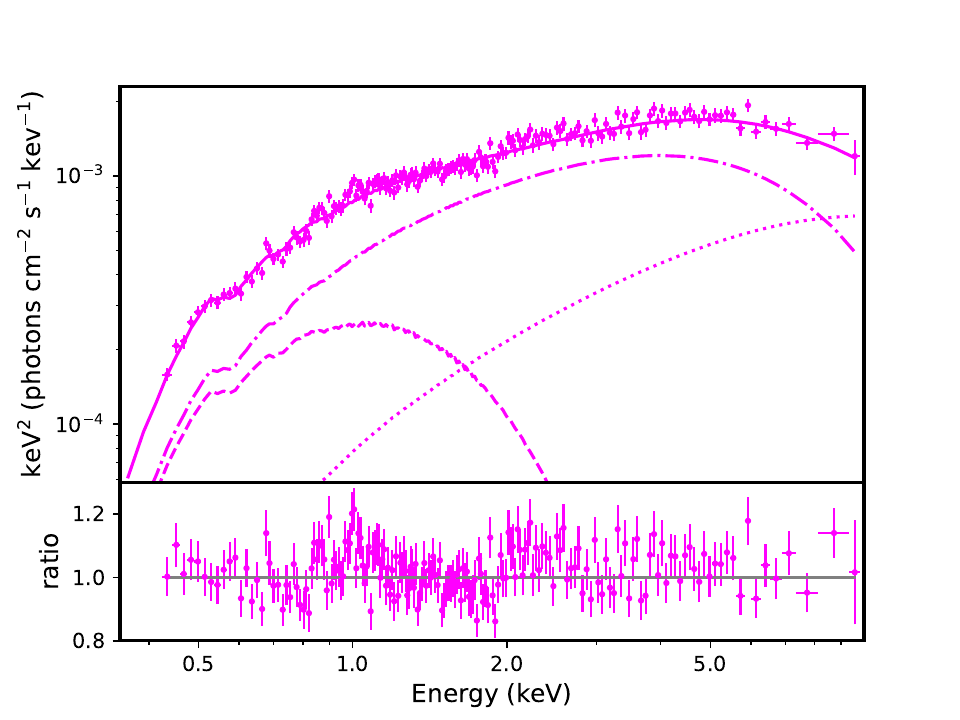}
   \caption{Combined EPIC-pn + MOS(1,2) spectra of the six spectral states extracted from the HID (see text and Fig.~\ref{spec_all}). { The colour convention is that indicated in Fig.~\ref{hid}}-{\it top}. The best fit is given by the {\sc tbabs(diskbb + diskPbb + cutoffpl)} model (see Table~\ref{table_fit}), { where the {\sc diskbb}, {\sc diskPbb} and {\sc cutoffpl} models are indicated with dashed, dot-dashed and dotted lines, respectively}. In the bottom panel { of each plot}, it is shown the ratio of data and model.}
   \label{spec}
\end{figure*}

\begin{figure*}
\center
\includegraphics[width=9.2cm]{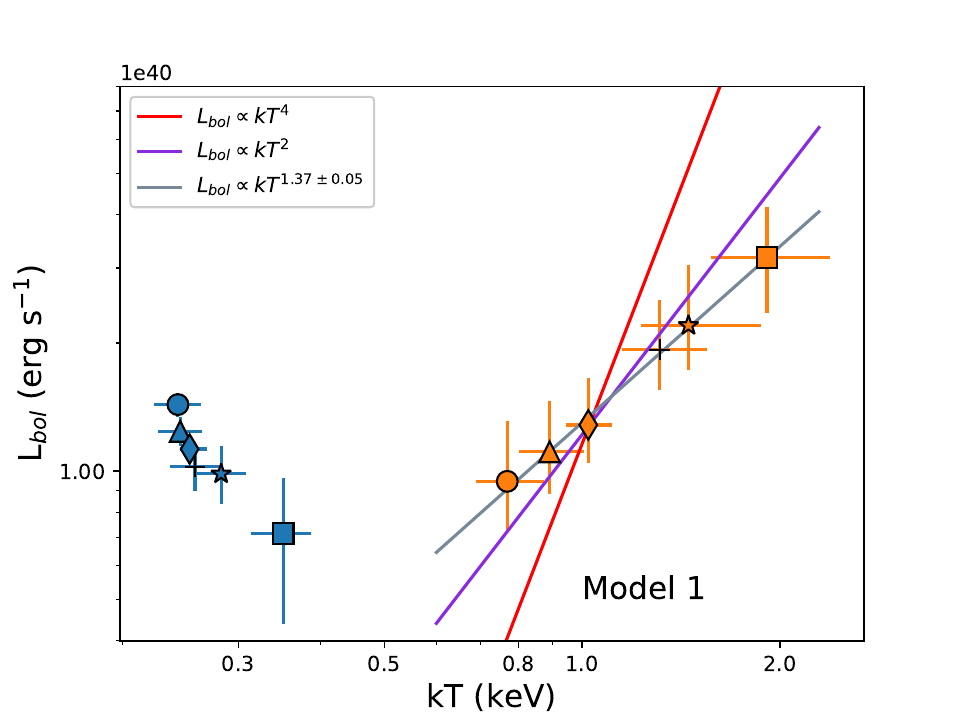}
 \hspace{-0.9cm}
\includegraphics[width=9.2cm]{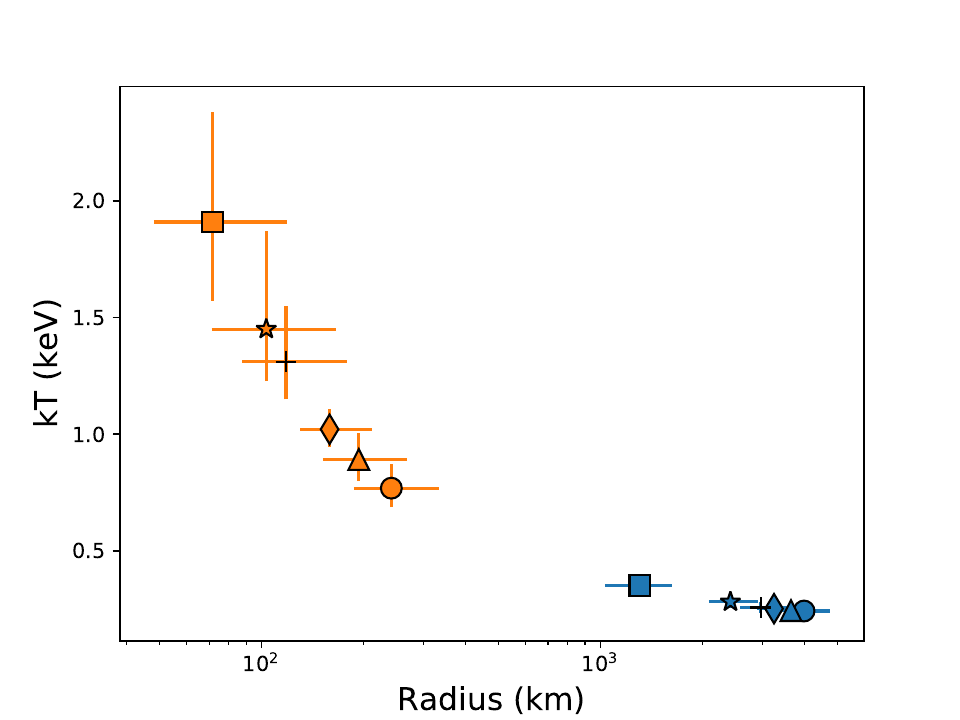}
   \caption{Thermal component unabsorbed bolometric luminosities (0.001--30 keV) versus the soft (blue points) and hard (orange points) disc temperature ({\it left}), and thermal component temperatures (soft disc in blue, hard disc in orange) vs their inner disc radii ({\it right}), for Model 1. We assumed an inclination of the system of 10 degrees. { The circle, triangle, diamond, cross, star and square points indicate the spectra from 1 to 6 indicated in Table~\ref{6states_tab}, respectively.}    
   }
   \label{lt4}
\end{figure*}

\begin{figure}
\center
\includegraphics[width=9.cm]{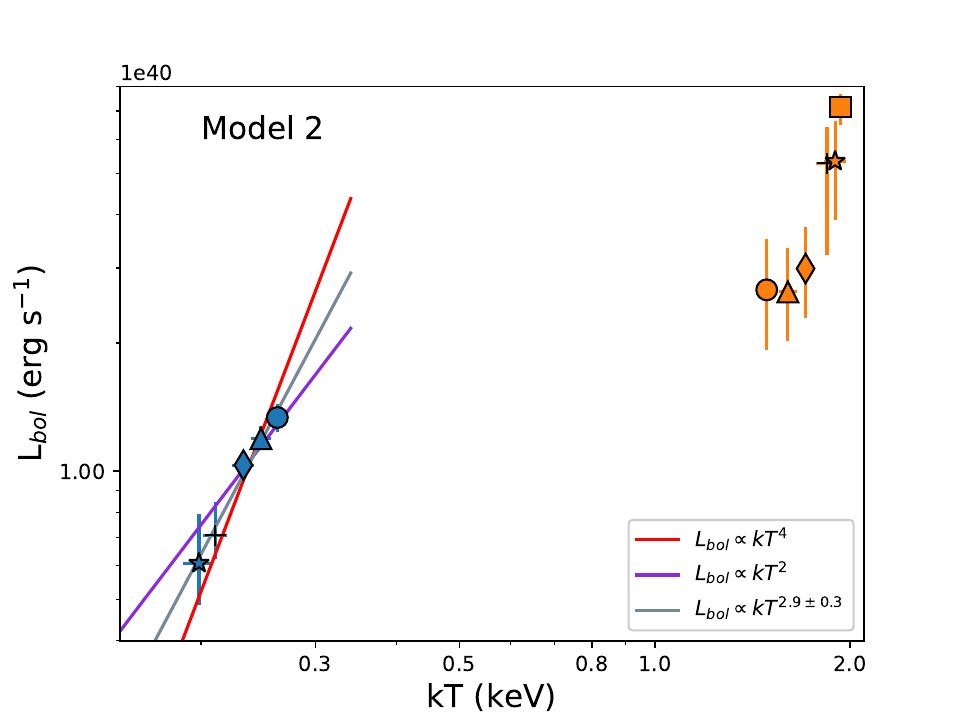}
   \caption{Thermal component unabsorbed bolometric luminosities (0.001--30 keV) versus the soft (blue points) and hard (orange points) disc temperature, for Model 2. { The circle, triangle, diamond, cross, star and square points indicate the spectra from 1 to 6 indicated in Table~\ref{6states_tab}, respectively.} }
\label{rt0.5}
\end{figure}

\subsubsection{Model 1: {\sc diskbb+diskPbb+cutoffpl}}

Since we found no significant variability between the individual spectra, we tied the column density (modelled with {\sc tbabs}) and the $p$ parameter of the {\sc diskPbb} amongst all the spectra. The {\sc cutoffpl} parameters were instead frozen to $\Gamma=0.59$ and E$_{\, cut}=7$ keV (i.e. the average of the cut-off power law parameters of the PULXs; \citealt{walton18}). We left free to vary all the remaining spectral parameters of the two thermal components and the normalizations of the {\sc cutoffpl} model. The best-fit gives $\chi^2_{\nu}\sim1.06$ for 3776 dof (null hypothesis probability (NHP) = $4\times10^{-3}$) { and we present the results in Table~\ref{table_fit}}). 
Although it is formally not statistically acceptable, we note that the continuum is generally well modelled and no evident residuals pop up, with the exception of a few narrow residuals around { 1--2} keV (see Fig. \ref{spec}).

{ The disc temperature and its bolometric luminosities are often compared in order to put constraints on the nature and geometry of the accretion disc itself. Indeed, for a standard, thin accretion disc, the luminosity and the inner disc temperature are expected to correlate as $L_{x} \propto T^4$ \citep{shakura73}, for a constant emitting area (i.e. constant inner disc radius). Instead, in the case of an advection dominated disc, the correlation changes as $L_{x} \propto T^2$ \citep{watarai01}. Therefore, looking for such correlations can give rough hints on the structure of the disc and its accretion regime. A number of works investigated the existence of these correlations in many ULXs \citep[e.g.][]{feng06, kajava09, pintore12, walton13, walton20, robba21, barra24} which resulted in a variety of relations, although we note that the outcomes are often dependent on the continuum spectral modeling. However, the search for correlations between luminosity and disc temperature is nevertheless a useful tool to characterise the accretion flow. Hereafter, we look for correlations for the thermal components used to model the spectral continuum of NGC 4559 X7.}

The soft thermal component slightly changes in flux but does not show significant variations of its temperature, apart for the highest flux state where such a component is very weak and poorly constrained ($\text{kT}_{\text{cold}} \sim 0.35$ keV). We found a tentative $L_{\text{cold}} \sim \text{kT}_{\text{in}}^{-2}$ anti-correlation but the uncertainties are large in both normalization and index, therefore we caution about it.  The harder thermal component has a temperature higher than $\sim1$ keV and it shows instead a correlation that is compatible with $L_{\text{hot}} \sim \text{kT}_{\text{in}}^{1.37\pm0.05}$ (Fig.~\ref{lt4}-left). The apparent radii span the ranges $\sim 1000~\text{km} - 4500$ km and $\sim 70~\text{km} - 250$ km for the soft and hard thermal component, respectively (Fig.~\ref{lt4}-right), assuming a low inclination of the system (10$^{\circ}$). 
We found that the apparent radii are not constant in time but that can vary with the source flux. The $L_x \propto kT^4$ correlation is given assuming the disc at a fixed inner radius \citep[e.g.][]{kubota04,done06}, hence the correlation found between luminosity and temperature can be strongly affected by the radius change. We highlight that the XMM0 data, fitted with the same model, are located in the L$_x$--kT region which is unrelated to the others. It indicates that the flux changes are key to the source evolution.

Notably, the cut-off power law still dominates the very high energies and its emission correlates with the global flux. We estimated that the flux ratio between the hard components (hard thermal disc plus cut-off power law) and the soft thermal components increases from 1 to $\sim$6, moving through the lowest to the highest flux spectral state, respectively. In particular, we found that the soft component manifests a maximum variation of a factor of $\sim1.5$ (taking into account the uncertainties), compared with the factor of $\sim3.5$ of the hard ones, with the latter being the main responsible for the spectral changes (as seen also from the lightcurves in Fig.~\ref{lightcurve}). We further note that the {\sc diskPbb} luminosity and cut-off powerlaw luminosity are strongly correlated, suggesting that their spectral changes are triggered by the same physical process.

\subsubsection{Model 2: {\sc diskbb+diskPbb+cutoffpl} with a constant radius}

The previous model gave a complex evolution of the spectral shape of X7 and indicated that the emitting radii of the thermal components change during time, although by no more than a factor of 2-3 taking into account the quite large uncertainties.
Therefore, we finally tested a model where the thermal component normalizations of both the {\sc diskbb} and {\sc diskPbb} are tied between the spectra (i.e. assuming fixed emitting radii). The best fit ($\chi^2_{\nu}\sim1.1$ for 3789 dof) { is shown in Table~\ref{table_fit_linked}. It} converged towards radii of $\sim3500$~km and $\sim50$~km for the soft and hard component, respectively. { Remarkably, the soft component shows a robust correlation between the inner temperature and its bolometric luminosity ($L_{\text{cold}} \sim \text{kT}_{\text{in}}^{2.9\pm0.3}$; Fig.~\ref{rt0.5}), which is neither compatible with a standard accretion disc nor with a fully advection dominated disc.} 
We note that the temperature of the cold component is at the highest during the minima of the global source flux and the component is negligible during the peaks of the flares, i.e. the component anti-correlates with the X7 bolometric flux. On the contrary, the behaviour of the hard component is less constrained ($L_{\text{hot}} \sim \text{kT}_{\text{in}}^{4.6\pm0.9}$; Fig.~\ref{rt0.5}), and compatible with a standard disc although the $p$ parameter of the {\sc diskPbb} indicates an advection dominated disc ($p\sim0.5-0.6$). These contradictory results might suggest that the assumption of a fixed inner radius might be suitable for the cold component but not for the hot one.
We also highlight that the cut-off power law component is negligible at the lowest X7 fluxes ($\ll 1\%$ of the total flux), while it is quite strong during the flares ($\sim25\%$).

\begin{table*}
\center
\begin{tabular}{llllllll}
\hline
Model 1 & Component & \#1 & \#2 & \#3 & \#4 & \#5 & \#6 \\
\hline
{\sc TBabs} & N$_{\text{H}}$ ($10^{22}$ cm$^{-2}$)  &  \multicolumn{6}{c}{${0.13}_{-0.01}^{+0.01}$} \\ 
\\
  \multirow{2}{*}{{\sc diskbb}} & kT$_{\text{in}}$ (keV)   & ${0.25}_{-0.02}^{+0.02}$ & ${0.24}_{-0.02}^{+0.02}$ & ${0.252}_{-0.008}^{+0.016}$ & ${0.26}_{-0.02}^{+0.02}$ & ${0.28}_{-0.03}^{+0.03}$ & ${0.35}_{-0.04}^{+0.04}$ \\ 
   & norm  & ${28}_{-7}^{+11}$ & ${23}_{-6}^{+9}$ & ${18}_{-4}^{+6}$ & ${15}_{-4}^{+6}$ & ${10}_{-3}^{+4}$ & ${3}_{-1}^{+1}$ \\ 
   \\
  \multirow{3}{*}{{\sc diskpbb}} & kT$_{\text{in}}$ (keV)   & ${0.77}_{-0.08}^{+0.10}$ & ${0.89}_{-0.09}^{+0.11}$ & ${1.02}_{-0.08}^{+0.09}$ & ${1.3}_{-0.2}^{+0.2}$ & ${1.5}_{-0.2}^{+0.4}$ & ${1.9}_{-0.3}^{+0.5}$ \\ 
   & p  & \multicolumn{6}{c}{${0.63}_{-0.04}^{+0.08}$} \\ 
  & norm  & ${0.10}_{-0.05}^{+0.08}$ & ${0.07}_{-0.03}^{+0.05}$ & ${0.04}_{-0.02}^{+0.03}$ & ${0.02}_{-0.01}^{+0.02}$ & ${0.02}_{-0.01}^{+0.02}$ & ${0.009}_{-0.006}^{+0.012}$ \\ 
  \\
  {\sc cutoffpl} & norm ($10^{-5}$)  & ${2.9}_{-0.6}^{+0.5}$ & ${3.8}_{-0.9}^{+0.7}$ & ${5.0}_{-0.8}^{+0.7}$ & ${8}_{-3}^{+2}$ & ${15}_{-6}^{+3}$ & ${11}_{-11}^{+6}$ \\  
\\
  & f$_{X}^b$ & $4.18_{-0.12}^{+0.12}$ & $4.33_{-0.14}^{+0.14}$ & $4.67_{-0.12}^{+0.12}$ & $6.11_{-0.20}^{+0.20}$ & $8.06_{-0.26}^{+0.24}$ & $8.24_{-0.17}^{+0.27}$\\
  & f$_{cold}^b$ & $2.13_{-0.14}^{+0.14}$ & $1.85_{-0.15}^{+0.15}$ & $1.68_{-0.13}^{+0.14}$ & $1.53_{-0.18}^{+0.20}$ & $1.47_{-0.22}^{+0.24}$ & $1.1_{-0.4}^{+0.4}$  \\
  & f$_{hot}^b$ & $1.41_{-0.34}^{+0.54}$ & $1.66_{-0.34}^{+0.52}$ & $1.92_{-0.36}^{+0.55}$ & $2.87_{-0.56}^{+0.88}$ & $3.28_{-0.7}^{+1.3}$ & $4.74_{-1.2}^{+1.5}$ \\

\\
\hline
 & $\chi^2/dof$ & \multicolumn{6}{c}{4011.87/3776}  \\
 \hline
\end{tabular}
\caption{Best-fit spectral parameters for the {\sc tbabs(diskbb + diskPbb + cutoffpl)} model (Model 1). Errors are at 90$\%$ uncertainty for each parameter of interest. \\
$^b$ Unabsorbed fluxes of the full model (f$_{X}$), cold temperature (f$_{cold}$) and hard (f$_{hot}$) temperature thermal models, respectively, in the 0.001--30 keV energy band in units of $10^{-12}$ erg cm$^{-2}$ s$^{-1}$.}
\label{table_fit}
\end{table*}

\begin{table*}
\center
\begin{tabular}{llllllll}
\hline
Model 2 & Component & \#1 & \#2 & \#3 & \#4 & \#5 & \#6 \\
\hline {\sc TBabs} & N$_{\text{H}}$ ($10^{22}$ cm$^{-2}$)  & \multicolumn{6}{c}{${0.157}_{-0.007}^{+0.007}$} \\ 
\\
  \multirow{2}{*}{{\sc diskbb}} & kT$_{\text{in}}$ (keV) & ${0.262}_{-0.009}^{+0.009}$ & ${0.247}_{-0.009}^{+0.009}$ & ${0.232}_{-0.009}^{+0.009}$ & ${0.210}_{-0.009}^{+0.009}$ & ${0.20}_{-0.01}^{+0.01}$ & ${0.0}_{0}^{+0.8}$ \\ 
   & norm  & \multicolumn{6}{c}{${22}_{-3}^{+4}$} \\ 
   \\
  \multirow{2}{*}{{\sc diskpbb}} & kT$_{\text{in}}$ (keV)   & ${1.49}_{-0.05}^{+0.05}$ & ${1.60}_{-0.05}^{+0.05}$ & ${1.71}_{-0.05}^{+0.06}$ & ${1.84}_{-0.07}^{+0.07}$ & ${1.90}_{-0.07}^{+0.07}$ & ${1.93}_{-0.07}^{+0.07}$ \\ 
   & p  & \multicolumn{6}{c}{${0.535}_{-0.008}^{+0.008}$} \\ 
   & norm  & \multicolumn{6}{c}{${0.0040}_{-0.0006}^{+0.0007}$} \\ 
   \\
  {\sc cutoffpl} & norm ($10^{-5}$)  & $<100$ & $<100$ & $<100$ & ${4.4}_{-1.2}^{+1.1}$ & ${12.6}_{-1.5}^{+1.5}$ & ${16.8}_{-1.8}^{+1.7}$ \\ 
\\

& f$_{X}^b$ & $6.18_{-0.16}^{+0.17}$ & $6.85_{-0.19}^{+0.20}$ & $7.72_{-0.13}^{+0.13}$ & $10.57_{-0.13}^{+0.13}$ & $13.24_{-0.12}^{+0.13}$ & $14.20_{-0.13}^{+0.13}$ \\

& f$_{cold}^b$ & $1.99_{-0.15}^{+0.15}$ & $1.78_{-0.07}^{+0.12}$ & $1.54_{-0.10}^{+0.10}$ & $1.05_{-0.13}^{+0.21}$ & $0.91_{-0.18}^{+0.27}$ & $<0.01$ \\

& f$_{hot}^b$ & $3.97_{-1.11}^{+1.25}$ & $3.93_{-0.92}^{+1.05}$ & $4.46_{-1.04}^{+1.13}$ & $7.87_{-3.1}^{+1.7}$ & $7.96_{-2.17}^{+1.96}$ & $10.67_{-0.99}^{+0.78}$  \\

\\
\hline
 & $\chi^2/dof$ & \multicolumn{6}{c}{4158.66/3789}  \\
 \hline
\end{tabular}
\caption{Best-fit spectral parameters for the {\sc tbabs(diskbb + diskPbb + cutoffpl)} model, when linking the normalization parameters of the thermal components for all the spectra (Model 2). Errors are at 90$\%$ uncertainty for each parameter of interest. \\
$^b$ Unabsorbed fluxes of the full model (f$_{X}$), cold temperature (f$_{cold}$) and hard (f$_{hot}$) temperature thermal models, respectively, in the 0.001--30 keV energy band in units of $10^{-12}$ erg cm$^{-2}$ s$^{-1}$.
}
\label{table_fit_linked}
\end{table*}

\bigskip
For completeness of our spectral analysis and in addition to the models described above, we also tested an absorbed {\sc diskbb+diskPbb}, {\sc diskPbb+diskPbb+cutoffpl} and a {\sc diskPbb+cutoffpl}, but they gave poorer or less constrained results. For this reason, they are not taken into account in this work. 

\subsection{High-resolution spectroscopy}
\label{High-res}

\begin{figure}
\center
\includegraphics[angle=0, width=8.4cm]{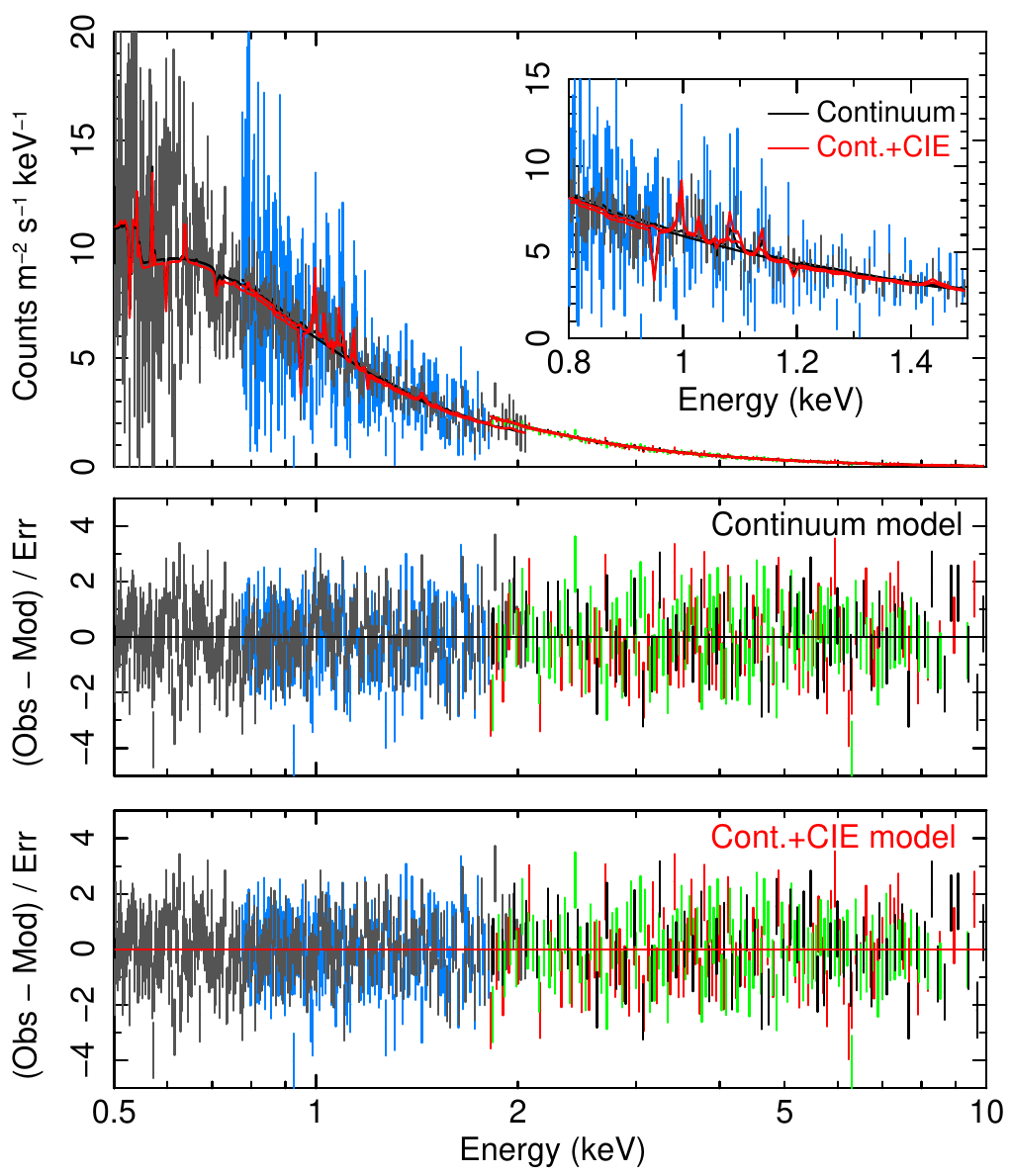}
   \caption{Combined RGS $1^{\rm st}$ and $2^{\rm nd}$ order (grey and blue), EPIC-pn and MOS 1,2 (black, red and green) spectra (top panel). The black and red lines refer to the continuum and continuum + CIE models, respectively. The corresponding residuals are shown in the middle and bottom panels. Note the Ne K - Fe L emission complex near 1 keV.}
   \label{spex_bestfit}
\end{figure}

We found that some residuals appear around 1 keV in the EPIC spectra, especially for the highest flux state (purple and cyan spectra in Fig.~\ref{spec}) {and more evident in spectra extracted from the entire observation (see \citealt{pintore21} for the archival data)}. Such features are common in ULX CCD spectra taken with \chandra\ and \xmm\ satellites \citep[e.g.][]{middleton14}. The combined RGS spectra along with the corresponding EPIC-pn and MOS 1, 2 spectra are shown in Fig. \ref{spex_bestfit}. {The broad feature seen in the EPIC data around 1 keV is resolved into a few emission lines} in the RGS spectra, which provide a high resolving power $R_{\, \rm RGS} = E / \Delta E \sim$ 200 (400) at 1 keV for the $1^{\rm st}$ ($2^{\rm nd}$) spectral order. Further, weaker, emission features seem to be present. {We focussed on the RGS data obtained by stacking the two longest observations (XMM1 and XMM2 as mentioned in Sect. \ref{data_reduction}). Besides, XMM0 has a different and fainter spectrum which makes it more prone to issues with the background subtraction.} 

Following previous works (see, e.g., \citealt{pinto21}), we kept the EPIC data only between 2--10 keV, i.e. outside the RGS energy, owing to the much poorer resolution ($R_{\, \rm EPIC} = E / \Delta E \lesssim 20$ below 2 keV, which is not sufficient to resolve narrow lines).  This avoids introducing further degeneracy between different models {which would provide comparable spectral fits for the EPIC data (its count rate higher than the RGS would basically lead the fit)}. The energy range between 0.5-2 keV (0.8-1.8 keV) was chosen for the RGS $1^{\rm st}$ ($2^{\rm nd}$) order because outside this range the spectrum is background dominated.

The RGS+EPIC spectral fits were performed with the {\sc SPEX} fitting package v3.07 (\citealt{Kaastra1996}), which is equipped with up-to-date atomic databases and flexible plasma models. The spectra were grouped according to the optimal binning discussed in \citet{Kaastra2016} and the Cash statistics ($C$-stat; \citealt{Cash1979}) was used. This has the advantage of smoothing the background spectra in the energy range with low statistics and removing narrow spurious features. We note, however, that further tests that were performed by rebinning to at least 25 counts per bin provided consistent results.

Following the modelling of the HID-selected EPIC spectra, we adopted a broadband continuum model consisting of two thermal components (disc-blackbody, {\sc dbb}\footnote{Note that {\sc SPEX} does not have the corresponding model of the {\sc diskPbb} in {\sc Xspec}} in {\sc SPEX}.) and a power law {with a high-energy cut-off (i.e. a {\sc pow} component multiplied for an {\sc etau} component with $\Gamma_{\rm pow}=0.59$ and E$_{\, cut}=7$ keV, respectively)}. The continuum model was self-consistently fit to the five averaged RGS+EPIC spectra except for a multiplicative constant that was kept free to vary with respect to the EPIC-pn (which was fixed to 1) to account for the typical 5\% cross-calibration uncertainties. The RGS+EPIC continuum model provides an overall $C$-stat of 811 (and a comparable {value of} $\chi^2$) for 659 dof. The continuum model is represented by the solid black line in Fig. \ref{spex_bestfit} with the corresponding residuals in the middle panel. The precise form of the continuum model has little impact on the modelling of narrow emission feature (at least in the case of collisionally-ionised plasma, see below).

\subsubsection{Gaussian line scan}
\label{Gaussian line scan}

Some narrow features and residuals can be seen in Fig. \ref{spex_bestfit} although they are clearly weaker than in other ULXs exhibiting softer X-ray spectra \citep[e.g.][]{pinto17}; this is not surprising given that, in hard ultraluminous X-ray sources \citep{sutton13}, there is an overwhelming continuum from the innermost regions (see e.g. \citealt{pinto17,kosec21}). For a more efficient visualisation and a clearer detection of narrow lines, we performed a standard scan of the spectra with a moving Gaussian line following the approach used in \citet{pinto16}. We tested five logarithmic grids 
with energies between 0.5 and 10 keV and different sizes according to the adopted line widths (300, 500, 1000, 1200 and 2000 points for a line FWHM of 10000, 5000, 2500, 1000 and 100 km s$^{-1}$, respectively). At each energy, the $\Delta C$ improvement to the best-fit continuum model is recorded. Its square root provides the single-trial significance; to distinguish between emission and absorption lines, we multiplied the $\sqrt{\Delta C}$ by the sign of the Gaussian normalisation.

\begin{figure}
\center
\includegraphics[width=8.4cm]{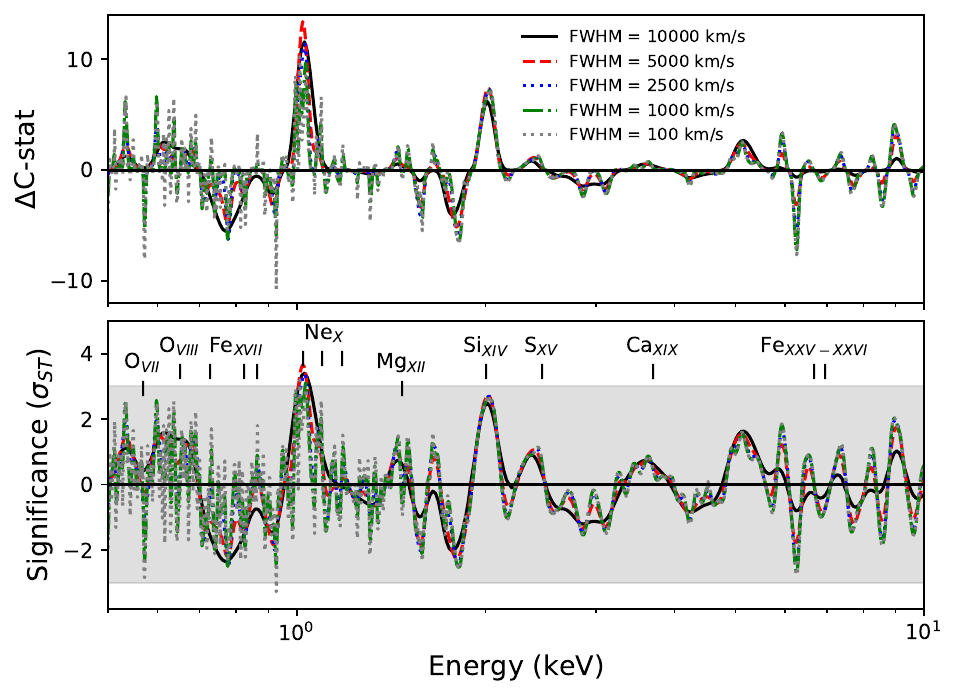}
   \caption{Gaussian line scan for the combined RGS (0.4-2 keV) and EPIC (2-10 keV) spectra and different line widths. Rest-frame centroids of the typical, dominant H- / He-like transitions are labelled.}
   \label{spex_gauscan}
\end{figure}

The results of the line scan obtained for the combined RGS+EPIC spectra are shown in Fig.\,\ref{spex_gauscan}. We remind the reader that in the $0.5-2$ keV energy band, only RGS data is considered. The line scan picked out the strong emission feature near 1.0 keV, which is broadly consistent with the Ne\,{\scriptsize X} rest-frame laboratory energy and some highly-ionised iron species (Fe\,{\scriptsize XX-XXIV}). Further, weaker, emission features are close to the O\,{\scriptsize VIII} (0.65 keV) and Si\,{\scriptsize XIV} (2.01 keV) lines. The absence of {lower-ionisation ionic species} such as Ne\,{\scriptsize IX} and O\,{\scriptsize VII}, frequently found in the RGS spectra of softer ULXs, would suggest that the plasma is either highly ionised or mainly in collisional-ionisation equilibrium (CIE). Some absorption features, albeit weaker, appear between $0.6-0.8$ keV, which mimic those previously found in many ULXs {(see Sect.~\ref{Physical models scan} and \ref{Discussion} for more detail).} The RGS line search for the broad 10,000 km s$^{-1}$ exhibit an overall shape {which mimics that of} the EPIC residuals (see Fig. \ref{spec}). 

\subsubsection{Physical models scan}
\label{Physical models scan}

In order to boost {the significance of the detection of the plasma component(s)} by simultaneously accounting for multiple lines, we performed a search throughout a multidimensional parameter space of plasma in {photo- (PIE) or collisional (CIE)} ionisation equilibrium. This technique prevents the fits from getting stuck in local minima, although it is computationally expensive (see, e.g., \citealt{kosec18a, pinto20}). Given the weakness of the absorption features we only tested PIE absorption models as the results would not be distinguishable from CIE absorption models. {Besides, absorption lines from CIE plasmas are not common in XRBs (see, e.g., \citealt{pinto21,Neilsen2023}).} We adopted Solar abundances {to save computation time}; this might have had an impact on the overall spectral improvement as the abundance pattern could differ from the Solar one {(see, e.g., \citealt{barra24} for the relevant case of ULX Holmberg II X-1}).

\begin{figure}
\center
\hspace{-0.8cm}
\includegraphics[width=8.5cm]{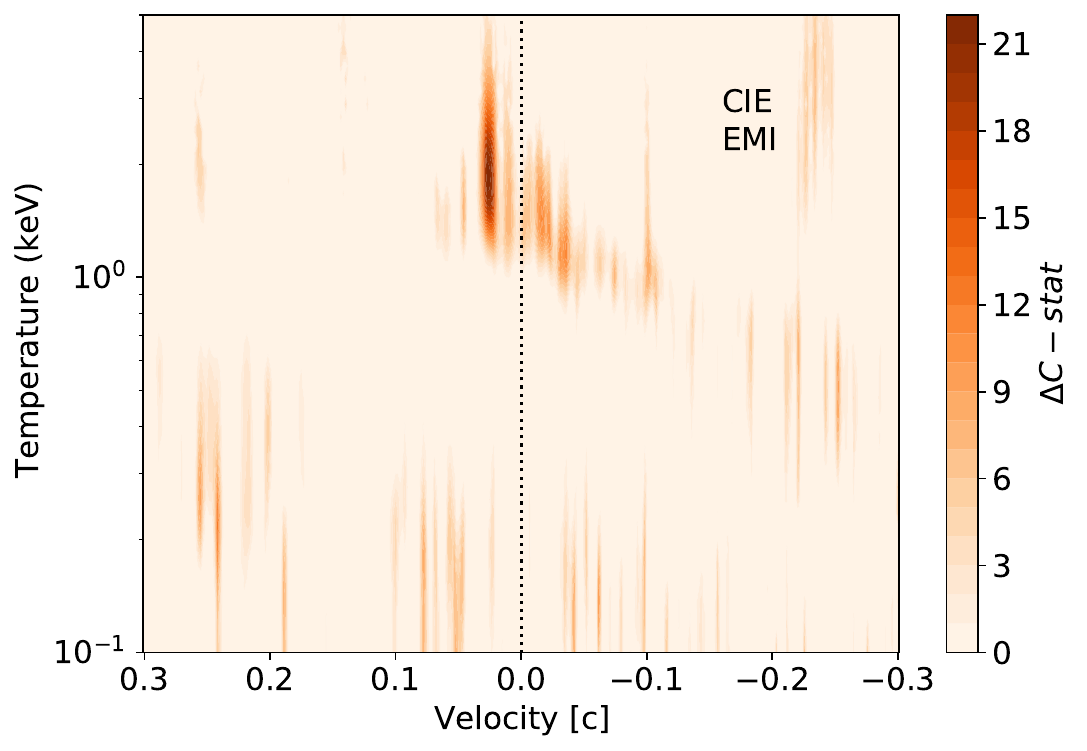}
\includegraphics[width=8.4cm]{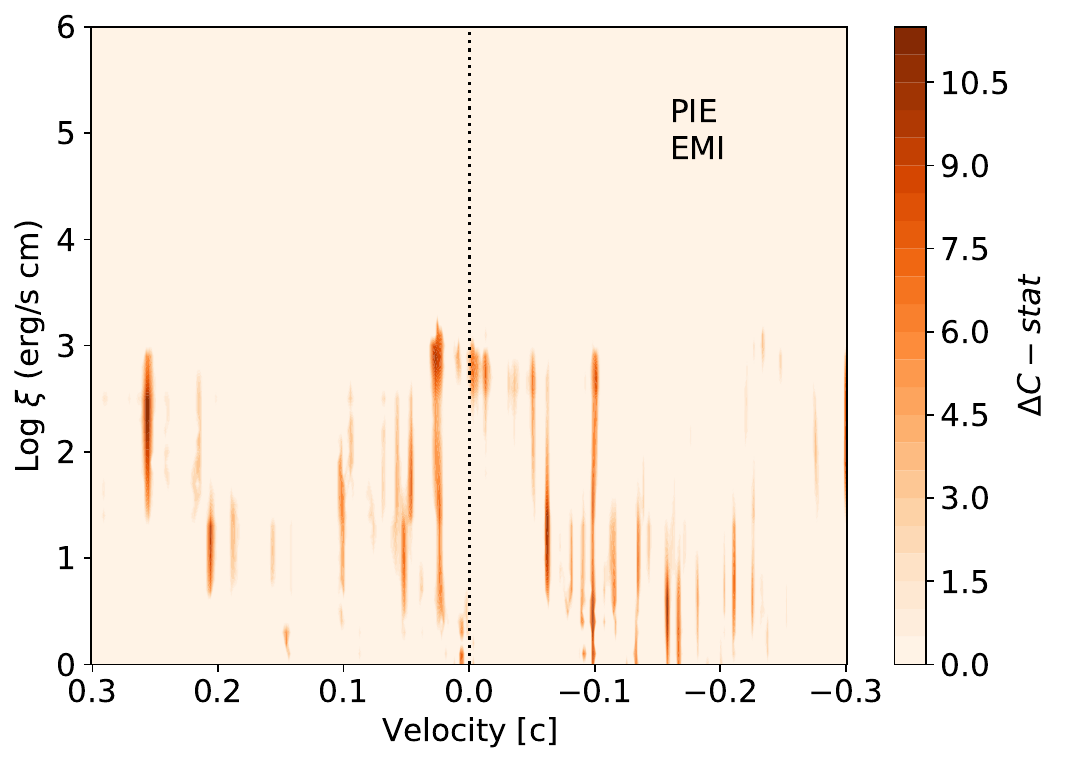}
\includegraphics[width=8.4cm]{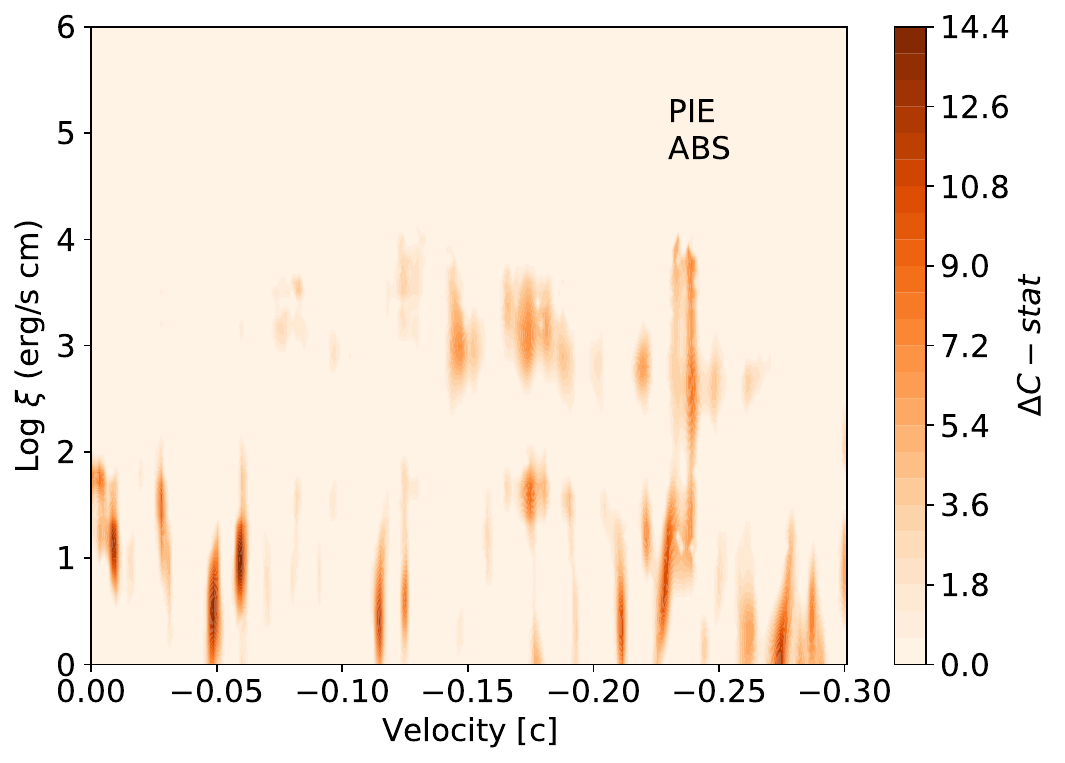}
   \caption{Top panel: multi-dimensional grids of emission models of plasma in collisionally-ionisation equilibrium for the combined RGS (0.4-2 keV) and EPIC (2-10 keV) spectra. Middle and bottom panels: grids of models of photoionised plasma in emission and absorption, respectively. Both scans adopted a line-broadening of 100 km s$^{-1}$ and Solar abundances. The best-fit CIE model is also shown in Fig. \ref{spex_bestfit}.}
   \label{spex_modscan}
\end{figure}

{\bf CIE line emission --} We first focused on the emission features by running a multidimensional automated scan with an emission model that assumes collisional ionisation equilibrium ({\sc cie} model in {\sc SPEX}). The {\sc cie} was multiplied by a redshift ({\sc reds} model) with {\sc flag=1} in order to test various line-of-sight velocities ($v_{\, \rm LOS}$).
We adopted a logarithmic grid of temperatures between 0.1 and 5 keV with 50 points. The range of $v_{\, \rm LOS}$ was chosen
between $-0.3c$ {(i.e. blueshift or outflowing plasma) and $+0.3c$ (redshift or inflowing plasma)}. The steps in velocities were selected according to the adopted velocity dispersion or line broadening ($v_{\, \rm step} =$ 500, 750, 1000, 1500 and 3000 km s$^{-1}$ for $v_{\, \rm RMS} =$ 100, 1000, 2500, 5000 and 10000 km s$^{-1}$, respectively). The {\sc cie} normalisation or emission measure was the only additional free parameter. Similarly to the gaussian scan, the $\Delta C$ improvement to the best-fit continuum model is recorded for each grid point and plotted in a k$T-v_{\, \rm LOS}$ graphic. The best results ($\Delta C_{\rm max}=21$) are obtained for $v_{\, \rm RMS}$ ranging from 100 to 1000 km s$^{-1}$ with the case for 100 km s$^{-1}$ shown in Fig. \ref{spex_modscan} (top panel). The peak {or best fit} is achieved at a temperature of 1.81 keV and a $v_{\, \rm LOS} = 0.0253\,c$ or 7600 km s$^{-1}$ (i.e. weakly redshifted) although a secondary, {less preferred}, solution is found around $-0.015\,c$. In Fig. \ref{spex_bestfit}, we also show the best-fit model, including the {\sc cie} component (solid red line in the top panel and corresponding residuals in the bottom panel). This component mainly accounts for the Ne K and Ne L lines around 1 keV. The {\sc cie} has a X-ray luminosity  $L_{\rm [0.3-10 keV]} \sim 1\times 10^{39}$ erg/s (for the adopted distance of 7.5\,Mpc). {The detection significance has to account for the look-elsewhere effect: in previous works, for RGS spectra with comparable statistics and with an identical parameter space explored, it was shown that a $\Delta C_{\rm max}=21$ corresponds to a conservative confidence level of about $3\,\sigma$ (see, e.g., \citealt{pinto17},\citealt{kosec18b}b, \citealt{kosec18a}a, \citealt{pinto20}). This has been further confirmed with an alternative method which uses cross-correlation instead of direct spectral fitting (see, e.g., \citealt{kosec21,xu22}).}

{\bf PIE line emission --} We also attempted to describe the emission lines with models of plasmas in photoionisation equilibrium. {\sc SPEX} is provided with the unique {\sc pion} model, which instantaneously calculates the photoionisation balance of the gas by using the fitted continuum model, produces the emission lines and recombination continua, and fits them to the spectra. Here, the balance is parameterised by the ionisation parameter, $\xi = L_{\rm ion} / ({n_{\rm H} \, R^2})$, where $L_{ion}$ is the ionising luminosity ({typically measured between 13.6 eV and 13.6 keV or 1--1000 Rydberg}), $n_{\rm H}$ is the hydrogen number density and $R$ is the distance from the ionising source. In analogy with the {\sc cie}, we created grids of {\sc pion} emission models adopting a logarithmic grid of ionisation parameters (log\,$\xi$ between 0 and 6 with steps of 0.1). For the $v_{\rm LOS}$ and $v_{\, \rm RMS}$ we used the same ranges tested over the {\sc cie} grids. The only free parameter for the {\sc pion} component is the column density, $N_{\rm H}$. To produce only emission lines, we chose a unitary solid angle $\Omega=1$ and a null covering fraction $f_{\rm cov}=0$. The {\sc pion} search resulted in a much lower improvement with respect to the {\sc cie} grids ($\Delta C_{\rm max}=11$ instead of 21) but also shows a peak at $v_{\, \rm LOS} = 0.0253\,c$ with log\,$\xi = 2.9$ (see Fig. \ref{spex_modscan}, middle panel).

{\bf PIE line absorption --} Finally, we also performed a scanning of grids through absorption models by selecting a solid angle $\Omega=0$ and a covering fraction $f_{\rm cov}=1$ for the {\sc pion} component. Since we do not expect redshifted absorption for the winds, we only selected a range of blueshifts for the LOS velocity, again following previous work. {\sc pion} in absorption works in the same way as in emission; the main difference is that the emission lines component is additive, while the absorption line component is multiplied for the continuum components. As for the emission, the only additional free parameter for each grid fit is the {\sc pion} $N_{\rm H}$. The absorption model grids provided worse results than the {\sc cie} grids (see Fig. \ref{spex_modscan}, bottom panel), which is expected given the weakness of the individual absorption features (see Fig. \ref{spex_gauscan}). The best fit is achieved for $v_{\, \rm LOS} = -0.06\,c$ and log\,$\xi = 1$, which would suggest a weak outflow, but the detection is not very significant given the maximum fit improvement of $\Delta C_{max}$ of just 14.

\section{Discussion}
\label{Discussion}

NGC 4559 X7 is a ULX with structured temporal and spectral variability in the form of random flares, which appear to be all flat-topped, and a marked spectral variability characterised by ``islands'' in the hardness-intensity diagram. These properties, not very common in the ULX population, make \src\ an intriguing case of study. 

\subsection{Accretion rate variability or accretion geometry changes?}
During the almost 20 years of X-ray observations, the source showed a significant and prolonged variability on both short (hours to days) and long-term (months to years) timescales. The long-term variability is characterized by a flux evolution of at least a factor of 5, with rapid increases and subsequent decays on timescales of a few weeks. We estimated that the peak X-ray luminosity is $\sim4\times10^{40}$ erg s$^{-1}$, while the average one is $\sim5\times10^{39}$ erg s$^{-1}$, for a distance of 7.5 Mpc.
The nature of such a variability might be explained either with changes in the accretion rate or in the accretion geometry or a combination of the two. 

The source emission shows at least three spectral states, two characterised by similar shapes but different fluxes (that we labelled as `persistent' states) and one (the `flaring' state) very hard and brighter than the others. There is now an increasing group of ULXs which present these `island` spectral states. Of this group, we mention the source M81 X-6, which was observed mainly in two spectral states with a characteristic harder-when-brighter behaviour \citep{weng18,gurpide21} and it was suggested to host a weakly magnetised NS \citep{amato23}. Recently, a long-term spectral evolution was investigated for Holmberg II X-1 \citep{gurpide21, barra24} and it was reported that the source can switch between hard and soft states directly related to the source luminosity. A possible explanation could be linked to a changing view of the inner regions of the accretion flow, shielded from time to time by an optically thick outflow ejected by the accretion disc. This is regulated by the closing of the funnel created by the outflow for increments in the accretion rate. For Holmberg II X-1, a super-soft state was also observed, while for NGC 4559 X7 such spectral regime was never seen. On the other hand, we cannot exclude that the sources contain the same compact object and slightly different accretion rates or inclination angle of the binary system.

In all the three states of \src\, the soft, thermal component does not change significantly in temperature but only in normalisation, of a factor of $\sim$2-2.5 (this work and Fig.~9 in \citealt{pintore21}). This suggests little accretion rate variability at large radii from the compact object. Instead, the hard component is very variable (more than a factor of 5) and it is responsible for the ``harder-when-brighter`` behaviour in X7, indicating that either the accretion onto the compact object is variable or other processes are in act so to affect its emission.

We can hypothesise that the global variability may be a combination of small fluctuations in the accretion rate and changes in the accretion geometry that outshines the inner regions emission, or as occultation from an outer, optically thick wind. Some evidence for winds in this source were found in the \xmm\ data. Both EPIC and RGS (XMM1 and XMM2) show spectral features around 1 keV which in RGS are resolved into a few Ne K and Fe L emission lines. A few absorption features may also be present, although the significance of them is not particularly high. The physical modelling favours collisional ionisation or shocks for the emission lines with a low velocity (redshift) of a few thousand km s$^{-1}$, which basically means plasma close to rest, given the limited resolving power and uncertainties around 1 keV. This is in line with the expectations for an intermediate-hard spectrum ULX where the binary system is expected to be seen close to face-on \citep{middleton15a}. In such a condition, winds are less prominent, especially in the case of a NS with high magnetic field as compact object, which would inhibit the launch of powerful, fast winds if the disc is truncated above the spherization radius. However, in most cases absorption lines were detected in ULX spectra with a number of counts higher than those collected for X7. The line emission component has a X-ray luminosity $L_{\rm [LE, 0.3-10 keV]} \sim 1 \times 10^{39}$ erg s$^{-1}$, which is a factor of 5--10 higher  than in typical ULXs \citep{pinto23}. Instead, the overall contribution to the source X-ray luminosity is a few \% in line with other ULXs. This might suggest that either the adopted distance of 7.5\,Mpc is an over estimation (quite unlikely because it is well constrained) or that the underlying emission process is particularly efficient in this source. The CIE best-fit solution indicates a redshift of $0.0253c$ or 7600 km s$^{-1}$ (very similar to some observations of the precessing jet in SS 433 -\citealt{marshall13,medvedev2019} - and the outflow in NGC 247 ULX-1 - \citealt{pinto21}). If the compact object was a NS, it might imply emission from near the magnetosphere or leakage towards the NS where gravitational redshift starts to kick in (rather than from a shocked disc wind). This might explain the unusual luminosity of the line-emitting plasma. Deeper observations are needed to place stronger constraints on the X7 features.

\subsection{A tentative super-orbital variability}
A tentative periodicity of $\sim240$~d was found in the \swift\ monitoring for the years $\sim$2020--2022. If it was orbital, the system would be quite large, making hard to explain the extreme mass transfer rate that sustains the accretion onto the compact object (unless it was a NS). A Be-binary system, with high inclination and eccentricity, could also explain the flux peaks (entrance in the Be-disc, i.e. higher accretion) and the lower flux state (the exit from the disc, i.e. lower accretion). A more likely scenario is that the periodicity is super-orbital as often seen in long-term monitored ULXs \citep[e.g.][]{fuerst16,walton16, brightman19, vasilopoulos20, salvaggio22}.

However, we note that the periodicity was not confirmed in the years 2023--2024 and that red-noise could also affect the significance of the signal. Therefore, we cannot robustly confirm yet the existence of a super-orbital variability in X7. On the other hand, the long-term lightcurve has large gaps where the behaviour of the source was unknown and we cannot rule out that X7 experienced abrupt flux decreases/increases during them. This may be important as sources with clear super-orbital periods (for example the PULXs NGC 5907 X-1 and NGC 7793 P13; e.g. \citealt{fuerst21}; \citealt{fuerst23}) did not recover immediately the periodicity right after a flux switch-off. Furthermore, it is not possible to exclude that the periodicity changes during time, as seen also in the Galactic source Her X-1 \citep[e.g.][]{leahy10} or in the PULX M51 ULX-7 \citep{brightman20, brightman22}.

\subsection{The flaring activity}
On short timescales, NGC 4559 X7 experienced a flaring-like activity, observed only at the very highest fluxes, i.e. in two \xmm\ and one \nus\ observations \citep{pintore21}. It cannot be proven yet if such an activity is directly connected to the high flux states, as long and continuous exposure observations are missing at low fluxes. The \swift\ monitoring did not catch any significant evidence of a flare, although we remark that the XRT exposures are short (on average, $\sim$2~ks) and the flares present a variable duration. The rising and decay times are quite fast (about 1 ks) and the peaks can last from a few hundreds of seconds up to tens of ks. This makes the flaring activity unpredictable and it is more probable that the \swift\ monitoring can observe only either the peak or the low flux state rather than the rising/decay epochs of the flares. Indeed, we estimate that the time spent during the rising/decay is on average $\sim10\%$ of the two high flux \xmm\ observations (total exposure of 195~ks). However, we note that the HID of the \swift\ data shows a bunch of observations with (tentative, because of the large uncertainties) hard spectra at the lowest fluxes. Hence, although the current evidence is quite weak, it cannot be rule out that flares happen also at low fluxes.

A quite impressive behaviour of the flares is that they are all flat-topped at the peaks, suggesting that there is a common mechanism able to limit the emission above a certain luminosity. A clear spectral variability is observed during the flares, mainly driven by the hard emission ($>1$~keV). 
The spectral analysis of X7 revealed that each spectral state can be described with the combination of two thermal components (disc-like emission) plus a non-thermal one (a cut-off power law). As mentioned before, the softer, thermal component is marginally varying during time between the two highest flux states (no more than a factor of $\sim$2 in flux), suggesting that there are no robust changes in the accretion rate (or at least not on the observational timescales) at large distance from the compact object. The soft component is quite sensitive to the continuum adopted to model the spectra, and it might show, alternatively, a correlation or an anticorrelation between temperature and luminosity. We suggest that the outer accretion flow could be a thick disc distorted by the outflow or the standard, outer disc itself. 

The apparent size of such a component is quite similar with the spherization radius of a 10 M$_{\odot}$ BH accreting at $\sim10$ times the Eddington rate. Indeed, the spherisation radius is defined as: 

\begin{equation}
R_{sph} = 27/4 \cdot \dot{m} \cdot R_g 
\end{equation}

and for $R_g = 15~\text{km}$ and $\dot{m}=10$, the radius is $\sim10^3$~km. This value is even smaller if the compact object was a NS ($R_{sph} \sim 1400$~km for $\dot{m}=100$). We should keep in mind that a flux variability similar to that of X7 is observed during some regimes (i.e. the $\rho$ class) of the BH GRS~1915+105 \citep[e.g.][]{belloni00}. Therefore, we cannot exclude that X7 hosts a massive BH, in a scaled version of Galactic BH accretion regimes. 

The harder components are the most intriguing spectral features of X7. They become more and more dominant while the flux increases (about a factor of 5 in flux variability). We note that the X7 spectra pivot around 0.8--0.9 keV (i.e. the luminosity of the soft component is anti-correlated with that of the hard component), indicating that the spectral evolution cannot be explained in terms of a dipping activity but rather to, again, either a change of the accretion geometry of the flow in the innermost regions or a switch-off/switch-on only of the accretion rate on the compact object. Precession effects from the wind could explain the spectral variability \citep[e.g.][]{middleton19}, but it would be hard to reconcile it with the absence of a quasi-periodic variability of the flares. 
The thermal, harder component shows a correlation, irrespective of the adopted continuum, between temperature and luminosity ($L_{\text{hot}} \propto kT_{\text{in}}^{1-4.5}$). This may suggest that the inner flow is compatible with an accretion disc, whose proper nature (standard or advection dominated) cannot be robustly constrained, although the spectral fits support likely an advection dominated disc ($p$$\sim$0.6). The smallest apparent emitting radius of the hot thermal component was estimated around { 70~km, during the brightest spectral state. If it was the innermost stable orbit of a BH, this would imply a M$_{BH}\sim8$ M$_{\odot}$} (assuming a non-rotating BH and an disc inclination of $10\degree$). In order to give a better estimate, we take into account a colour-correction factor of $f_{col} = 1.7$ \citep[e.g.][]{shimura95} and a boundary condition factor $\xi = 0.4$ \citep[e.g.][]{kubota98} for the accretion disc: this would imply a { larger} apparent radius and the BH mass estimate will rescale to { $\sim9$ M$_{\odot}$}. { However, the assumption that links the apparent radius of the high energy thermal component to the innermost stable orbit could not be appropriate, therefore we still cannot rule out} the presence of a NS. We remark that the spectral modeling we adopted for NGC 4559 X7 is consistent with that used for the PULXs \citep[e.g.][]{walton18}.

\subsection{A new candidate PULX?}
It is worth to note that we found some evidence for a pulsed signal with a period of $P\sim2.6\text{s}-2.7\text{s}$. Although marginally significant in the single observations, it is noteworthy that similar signals were spotted in two different \xmm\ observations (the ones at the highest flux). The signals present the same parameter space for the orbital parameters and, in addition, our tests exclude that they are an artefact. These findings strengthen the existence of the pulsation in X7. Applying orbital corrections, we estimated that the orbital period falls in the range 2--4 days with a quite unconstrained $A_{x}~\text{sin}(i)$ ($\sim$20--120 lt-s) that, if confirmed, could support a super-orbital origin for the 240~d periodicity in X7. If the pulsed signal is real, the compact object would be unequivocally a NS accreting well above its Eddington limit (about a factor of 100). The orbital parameters are compatible with those observed in the PULXs \citep[e.g.][]{israel16a,fuerst16,rodriguez19,bachetti22, belfiore24}. On the other hand, we remark that our estimate of the significance may be affected by caveats, therefore we still keep caution on claiming the discovery of a new PULX before further confirmations.
However, from a spectral point of view, PULXs are generally very hard \citep[e.g.][]{pintore16,gurpide21} similarly to what spectrally found for NGC 4559 X7 in XMM1 and XMM2, where the source might show evidence for pulsations. Many hints are hence pointing towards a NS in the source.

We found that the spin period would be affected by a strong secular spin-down of $\dot{P}\sim10^{-9}$ s s$^{-1}$, which is very high and quite unexpected since the source was always observed in accretion. If the propeller\footnote{The effect that inhibits the accretion on the NS when the magnetospheric radius is larger than the corotation radius.} was active in between our observing epochs, we should have detected a luminosity jump of at least one order of magnitude for the observed spin period of 2.6~s, which is not present in the long-term lightcurve though. Indeed, the maximum jump is of about a factor of 5. However, it was proposed that the lack of advection in the accretion flow and a combination of magnetic dipole field and $\dot{m}$ could imply small drops in luminosity during propeller \citep{middleton23}. Furthermore, it cannot be forgotten that numerous gaps are present in the long-term lightcurve, and even though there are no real clues about any switch-off of the source during them, we cannot rule-out such an hypothesis.

A first-guess possibility that we might consider is that the largest part of the source flux does not come actually from the accreting matter falling into the compact object, but rather from the one beyond the magnetospheric radius. Indeed, if the source was continuously in a propeller phase (and the feature around 1 keV in the highest flux spectrum can be perhaps due to shocked gas produced by this effect), the vast majority of the emission would arise from the matter outside the magnetospheric radius. From time to time, there could be a leakage of material to the NS surface which is responsible of the hardening of the spectral emission, an increase of the flux during the flares and possibly the appearance of the pulsation. In such a scenario, the pulsation is expected to be intermittent but this cannot be proven with the current counting statistics. The accretion process might proceed through a different mechanism as well.

We estimate that the co-rotation radius for a NS spinning at 2.7s is $R_{\text{cor}} = (G~M_{NS}~P^2 / (4\pi^2))^{1/3} \sim 3300$~km (where G is the gravitational constant and $M_{NS}$ is the NS mass). This is roughly compatible with the spherization radius that we estimated above, which would explain the weak winds due to an intrinsic low efficiency of the radiation pressure \citep{mushtukov19}. In such a condition, the hard components (the thermal disc and the cut-off power law) may be the emission from an accretion curtain inside the magnetosphere and an accretion column above the NS.

If the source was at the spin equilibrium, the NS dipolar magnetic field would be very high ($\sim10^{13}$ G), truncating the disc likely before the spherization radius.
However, should massive outflows be confirmed in the future, this would likely imply in turn that the source has a smaller magnetic field and its accretion proceeds in absence of propeller. The secular spin down is therefore tricky to explain, unless we take into account physical mechanisms as e.g. the threading of the magnetic field at large radii, which exerts a net torque on the NS which is more effective than the spin-up torque due to the accretion. 
Recently, \citet{bachetti22} and \citet{liu24} reported that the PULX M82 X-2 experienced periods of spin-down while accreting, suggesting that the source is at the spin-equilibrium and small changes in the accretion rate can trigger the propeller. Although there is no indication of a spin-equilibrium scenario for X7 (see above), the spin-down in NGC 4559 X7 might be similar to M82 X-2. Another pulsar with a strong secular spin-down while accreting is the Galactic GX 1+4 \citep[see e.g.][ and reference therein]{gonzalezgalan12}, even though it is a low-mass X-ray binary probably unlike NGC 4559 X7 \citep{soria04} and the other PULXs. It is noteworthy to remark that the spin-down of GX 1+4 is even stronger ($\sim 10^{-7}$ s s$^{-1}$) than that of NGC 4559 X7. However, given the longer spin period of GX 1-4 than X7, the $\dot{P}/P$ for the two sources differs by no more than a factor of 3, which is quite impressive seen the profound differences in their properties. 4U 1538-522, a Galactic high-mass X-ray binary pulsar, shows also a clear spin-down without any flux variability \citep[e.g.][and references therein]{hu24}. LMC X-4 is also characterised by almost quasi-periodic variations of spin-up and spin-down episodes \citep[e.g.][]{molkov17}.

Some models proposed that the geometry of the interaction between magnetosphere and accretion disc has an elliptical shape \citep{perna06}, suggesting that spin-up and spin-down can act simultaneously in different positions of the magnetosphere or that a retrograde accretion disc could be responsible for the pulsar spin-down during accretion \citep[e.g.][]{makishima88, dotani89, nelson97}. However, the latter scenario is hard to reconcile with NGC 4559 X7 as it has to be explained how a retrograde disc can be sustained for many years.
Therefore, no clear explanations could be currently proposed to describe the source behaviour. 

\subsection{An exotic scenario: a ms pulsar in NGC 4559 X7.}

If the reported spin period for NGC 4559 X7 is merely a statistical fluctuation, we propose a further, exotic scenario for this source. The flat-topped flares suggest that a common mechanism inhibits the accretion above a certain level, decaying rapidly towards lower fluxes with softer spectral shapes. If the rapid passage from the flares to a persistent emission was due to a propeller switch-on/switch-off, the luminosity jump ($\Delta L \propto 170~P^{2/3}~M_{1.4M_{\text{sun}}}^{1/3}~R_{10^6cm}^{-1}$; e.g. \citealt{campana18}) of a factor of $\sim3$ (or 5 if we consider the lowest source flux ever observed) would imply a NS spinning at a few ms periods ($P = 2\text{ms}-5\text{ms}$). The dipolar Larmor luminosity \citep[see][for a review]{burderi13} is:

\begin{equation}
L_{Lar} = 3.85 \times 10^{35}~B_{8}^2~R_6^{6}~P_{-3}^{-4}~\text{erg}~ \text{s}^{-1},
\end{equation}

where B$_{8}$ is the dipolar magnetic field in units of $10^8$G, R$_{6}$ is the NS radius in units of $10^6$ cm and P$_{-3}$ is the spin period in units of 1 ms. This luminosity has a strong dependence on the spin period and the magnetic field. For a ms pulsar of a period in the range $P=2\text{ms}-10\text{ms}$, in order to have a luminosity significantly smaller than the one observed in NGC 4559 X7 (minimum luminosity of $\sim5 \times 10^{39}$ erg  s$^{-1}$) so that its contribution can be ignored, the dipolar magnetic field has to be $<10^{12}$ G. However, if the source is seen at small angles, i.e. face-on, and the walls of the outflows can beam the inner region emission, the isotropic source luminosity would be smaller (for example, a beaming factor of 0.1 implies an order of magnitude less) and therefore the magnetic field reduces consequently.

The co-rotation radius, defined as
 
\begin{equation}
R_{cor} = \left(\cfrac{G~M_{NS}~P^2}{4~\pi^2}\right)^{1/3},
\end{equation}

for a ms pulsar of 2ms--10ms period, is in the range 20--80 km, which is compatible with the apparent emitting radius of the hot accretion disc found from the spectral analysis (taking into account also the colour-correction and boundary condition factors). In the case of spin-equilibrium, to explain the switch-on/switch-off of the propeller, the magnetic field would be of the order of $10^{12}-10^{13}$~G, to have a luminosity limit \citep[e.g.][]{campana18} compatible with the X7 luminosity peak ($4\times10^{40}$ erg s$^{-1}$). Taking into account the beaming, the magnetic field will be lower than that. 

Currently, this exotic scenario cannot be proven because there are no available X-ray observations with a ms time-resolution and a large number of collected counts to measure very small pulse fractions. We suggest that \xmm\ observations in \textit{TIMING} mode of this and many more bright ULXs without pulsation detections will be necessary to find evidences of such a short NS period.

\section{Conclusions}
We investigated the spectral and temporal properties of the ULX NGC 4559 X7. Variability up to a factor of $\sim5$ on long-term timescales were observed thanks to a \swift\ monitoring. Also \xmm\ caught twice a rapid short-term variability, whose behaviour resembles a flaring activity. Flares are mainly driven by changes in the hard emission ($>1$ keV, while the softer emission is more stable in time (less than a factor of 2--2.5 in variability).

We analysed X7 spectra by adopting models based on the combination of two thermal components plus a cut-off power law, i.e. a spectral modeling often used for the PULXs. We found that such a continuum is suitable for the X7 spectra. The soft component is quite sensitive to the adopted continuum and it is not clear if its temperature and luminosity follow a correlation or an anti-correlation. On the other hand, the hard thermal component correlates in the two quantities, irrespective of the adopted continuum. Hence, we suggest that the soft component is associated to the outer regions of the disc or a weak outflow above it, while the hard thermal component is likely an advection dominated accretion disc. The apparent emitting radii of the two components are $\sim2000$~km and $\sim100$~km, respectively.

A pulsed signal at 2.6s--2.7s was found in the \xmm\ observations taken in 2019 and 2022. Although the signal is found in two observations separated by years and sharing the same parameter space for the orbital properties, on one hand it increases its significance while, on the other, the spin period variability implies a strong spin-down of $\sim10^{-9}$ s s$^{-1}$. The latter is hard to explain as X7 was observed always in an apparent accretion regime. Several hypotheses were taken into account, such as threading of the disc, retrograde accretion disc and high NS dipolar magnetic field, but none appears to be conclusive. New \xmm\ observations will be needed to definitely constrain the existence of such a pulsating signal.

\section*{Acknowledgements} 

This work has been partially supported by the ASI-INAF program I/004/11/4 and program I/004/11/6. FP, GR, CP, MDS, AD, AW acknowledge support from INAF grants OBIWAN (Observing high B-fIeld Whispers from Accreting Neutron stars, 1.05.23.05.12) and BLOSSOM. CP acknowledges support from PRIN MUR 2022 SEAWIND 2022Y2T94C funded by NextGenerationEU. MDS acknowledges support from the INAF grant ACE-BANANA (1.05.12.05.17).
DJW acknowledges support from the Science and Technology Facilities Council (STFC; grant code ST/Y001060/1). NOPV acknowledges support from INAF and CINECA for granting 125,000 core-hours on the Leonardo supercomputer to carry out the project ”Finding Pulsations with Evolutionary Algorithms".

\section*{Data Availability}
All of the data underlying this article are already publicly available from ESA's XMM-Newton Science Archive (https://www.cosmos.esa.int/web/xmm-newton/xsa), NASA's HEASARC archive (https://heasarc.gsfc.nasa.gov/), and the Chandra Data Archive (https://cxc.harvard.edu/cda/). The SPEX code is also publicly available at https://github.com/ciropinto1982.

\addcontentsline{toc}{section}{Bibliography}
\bibliographystyle{aa}
\bibliography{biblio}

\section{Appendix A}

PSO searches for an optimal set of orbital parameters that yield the highest signal power in this space, without relying on a pre-determined grid. It relies on the movement of "particles" (i.e., candidate solution in the search space), whose positions are updated iteratively. The position update relies on two formulas. The first defines how the velocity of a particle is updated:

\begin{equation}
    v(t+1) = \omega v(t) + c_pr_p(p - x(t)) + c_gr_g(g - x(t))
    \label{eq:velocity}
\end{equation}

The second defines how its position is updated based on the velocity:

\begin{equation}
    x(t+1) = x(t) + v(t)
    \label{eq:position}
\end{equation}

Here, $\omega$ is the inertia and controls the influence of the previous velocity $v(t)$ on the new velocity $v(t+1)$, $x$ is the particle position, $p$ is the best position reached by the current particle (i.e., the position with the highest power), $g$ is the best position reached by any particle, $r_p$ and $r_g$ are two random numbers, and $c_p$ and $c_g$ are acceleration coefficients and represent, respectively, the cognitive component (i.e., attraction towards $p$) and the social component (i.e., attraction towards $g$).

The algorithm consists of 4 steps:

\begin{itemize}
    \item Initialization: $n$ particles are randomly distributed in the search space with a random initial velocity
    \item Evaluation: the power associated with each particle is computed
    \item While a stopping criterion is not met:
    \begin{itemize}
        \item Update: the particles adjust their velocities and positions using Equations \ref{eq:velocity} and \ref{eq:position}
        \item Evaluation of the particles at the new positions
    \end{itemize}
\end{itemize}

In this work, the search space has been divided into six non-overlapping regions:

\begin{enumerate}
    \item $A_{x}~\text{sin}(i) \in [20, 40]$ and $P_{orb} \in [2, 3]$ 
    \item $A_{x}~\text{sin}(i) \in [40, 60]$ and $P_{orb} \in [2.5, 3.5]$
    \item $A_{x}~\text{sin}(i) \in [60, 80]$ and $P_{orb} \in [3, 4]$
    \item $A_{x}~\text{sin}(i) \in [80, 90]$ and $P_{orb} \in [3, 4]$
    \item $A_{x}~\text{sin}(i) \in [90, 110]$ and $P_{orb} \in [3.5, 4.0]$
    \item $A_{x}~\text{sin}(i) \in [110, 120]$ and $P_{orb} \in [3.5, 4.0]$
\end{enumerate}

For each, three searches were performed, in different frequency ranges: [0.333, 0.42], [0.366, 0.367] and [0.375, 0.385]. The $|\dot P/P|$ values range from $10^{-5}$ to $10^{-12}$.

Each search stops after 100,000 evaluations. However, if the current best power has been found in the last 1,000 evaluations, the algorithm extends the search by a additional  10,000 evaluations. This process continues in increments of 10,000 evaluations until no improvements are detected in the last 1,000 iterations. As for the algorithm hyperparameters, the number of particles is set to $n = 40$, the cognitive component is set to $c_p = 1.496172$, the social component is set to $c_g = 1.496172$ and the inertia is set to $\omega = 0.72984$, as in \citet{4223164}.

\begin{table}
\caption{Log of the observations used in this work.}
\scalebox{0.65}{
\label{list_of_log}
\begin{tabular}{@{\makebox[1.3em][l]{\rownumber\space}} | llccr}
\hline
Instr. & Obs.ID & Start & Stop & Exp. \\
 & & \multicolumn{2}{c}{[YYYY-MM-DD hh:mm:ss (TT)]} & [ks] 
\gdef\rownumber{\stepcounter{magicrownumbers}\arabic{magicrownumbers}} \\
\hline
\chandra & 2026 & 2001-01-14T16:24:39 & 2001-01-14T19:27:35 & 9.4 \\
\chandra & 2027 & 2001-06-04T01:14:29 & 2001-06-04T04:43:23 & 10.7 \\
\chandra & 2686 & 2002-03-14T05:13:59 & 2002-03-14T06:34:19 & 3.0 \\
\xmm & 0152170501 & 2003-05-27T03:07:30 & 2003-05-27T14:19:33 & 42.2 \\
\swift & 00032249001 & 2012-01-14T15:53:31 & 2012-01-14T20:41:16 & 3.9 \\
\swift & 00032249003 & 2012-01-16T01:01:11 & 2012-01-16T06:17:57 & 4.2 \\
\swift & 00032249004 & 2012-01-18T01:33:23 & 2012-01-18T19:21:55 & 3.1 \\
\swift & 00032249005 & 2012-01-20T04:35:58 & 2012-01-20T12:58:55 & 3.8 \\
\swift & 00576064000 & 2013-10-28T05:46:26 & 2013-10-28T07:38:28 & 3.5 \\
\swift & 00032249006 & 2014-02-27T18:39:59 & 2014-03-03T11:21:54 & 4.9 \\
\swift & 00032249007 & 2014-03-05T23:46:07 & 2014-03-05T23:59:55 & 0.8 \\
\xmm & 0842340201 & 2019-06-16T19:03:44 & 2019-06-17T15:06:27 & 74.3 \\
\swift & 00088825001 & 2019-06-17T22:40:58 & 2019-06-17T23:08:53 & 1.7 \\
\swift & 00032249008 & 2019-12-12T00:46:13 & 2019-12-12T01:09:52 & 1.4 \\
\swift & 00032249009 & 2019-12-19T12:50:14 & 2019-12-19T13:12:53 & 1.4 \\
\swift & 00032249011 & 2020-01-02T08:32:43 & 2020-01-02T08:47:53 & 0.9 \\
\swift & 00032249012 & 2020-01-09T06:04:44 & 2020-01-09T06:21:52 & 1.0 \\
\swift & 00032249013 & 2020-01-16T00:40:59 & 2020-01-16T00:50:53 & 0.6 \\
\swift & 00032249014 & 2020-01-23T15:43:19 & 2020-01-23T16:05:53 & 1.3 \\
\swift & 00032249015 & 2020-01-30T16:42:51 & 2020-01-30T17:04:52 & 1.3 \\
\swift & 00032249016 & 2020-02-06T12:51:41 & 2020-02-06T13:14:53 & 1.4 \\
\swift & 00032249017 & 2020-02-13T13:49:35 & 2020-02-13T14:08:54 & 1.2 \\
\swift & 00032249018 & 2020-02-20T00:24:36 & 2020-02-20T02:08:52 & 1.4 \\
\swift & 00032249019 & 2020-02-27T12:21:16 & 2020-02-27T12:43:52 & 1.4 \\
\swift & 00032249020 & 2020-03-05T11:39:38 & 2020-03-05T12:01:52 & 1.3 \\
\swift & 00032249021 & 2020-03-12T00:06:31 & 2020-03-12T00:26:52 & 1.2 \\
\swift & 00032249022 & 2020-03-19T10:17:34 & 2020-03-19T10:40:54 & 1.4 \\
\swift & 00032249023 & 2020-03-26T12:51:57 & 2020-03-26T13:14:54 & 1.4 \\
\swift & 00032249024 & 2020-04-02T12:11:47 & 2020-04-02T12:33:54 & 1.3 \\
\swift & 00032249025 & 2020-04-09T11:36:37 & 2020-04-09T11:53:52 & 1.0 \\
\swift & 00032249026 & 2020-04-16T07:50:00 & 2020-04-16T08:10:54 & 1.2 \\
\swift & 00032249027 & 2020-04-23T02:09:17 & 2020-04-23T02:32:54 & 1.4 \\
\swift & 00032249028 & 2020-04-30T11:11:28 & 2020-04-30T11:33:52 & 1.3 \\
\swift & 00032249029 & 2020-05-07T12:05:07 & 2020-05-07T12:28:54 & 1.4 \\
\swift & 00032249030 & 2020-05-14T11:29:05 & 2020-05-14T11:45:53 & 1.0 \\
\swift & 00032249031 & 2020-05-21T12:24:35 & 2020-05-21T12:42:53 & 1.1 \\
\swift & 00035479001 & 2006-01-02T18:56:44 & 2006-01-03T04:47:56 & 5.3 \\
\swift & 00032249032 & 2020-05-28T11:29:03 & 2020-05-28T11:51:52 & 1.4 \\
\swift & 00032249033 & 2020-06-04T07:44:35 & 2020-06-04T08:11:53 & 1.6 \\
\swift & 00032249034 & 2020-06-11T07:06:17 & 2020-06-11T07:28:53 & 1.4 \\
\swift & 00032249035 & 2020-06-18T04:46:38 & 2020-06-18T05:09:52 & 1.4 \\
\swift & 00032249036 & 2020-06-25T07:14:56 & 2020-06-25T07:38:53 & 1.4 \\
\swift & 00032249037 & 2020-07-02T05:00:02 & 2020-07-02T05:20:53 & 1.2 \\
\swift & 00032249038 & 2020-07-09T01:00:01 & 2020-07-09T01:23:53 & 1.4 \\
\swift & 00032249039 & 2020-07-16T03:41:41 & 2020-07-16T04:00:52 & 1.1 \\
\swift & 00032249040 & 2020-07-23T20:20:22 & 2020-07-23T20:46:51 & 1.6 \\
\swift & 00032249041 & 2020-07-30T07:01:53 & 2020-07-30T07:25:55 & 1.4 \\
\swift & 00032249042 & 2020-08-06T12:36:18 & 2020-08-06T13:01:52 & 1.5 \\
\swift & 00032249043 & 2020-10-28T13:48:32 & 2020-10-28T14:12:52 & 1.5 \\
\swift & 00032249044 & 2020-11-04T16:15:16 & 2020-11-04T16:37:53 & 1.4 \\
\swift & 00032249045 & 2020-11-11T05:57:36 & 2020-11-11T17:30:53 & 1.5 \\
\swift & 00032249046 & 2020-11-18T03:39:42 & 2020-11-18T04:01:53 & 1.3 \\
\swift & 00032249047 & 2020-11-25T06:09:00 & 2020-11-25T06:33:52 & 1.5 \\
\swift & 00032249048 & 2021-01-24T17:41:56 & 2021-01-24T18:09:53 & 1.7 \\
\swift & 00032249049 & 2021-01-31T09:03:13 & 2021-01-31T22:09:54 & 1.6 \\
\swift & 00032249050 & 2021-02-07T16:35:44 & 2021-02-07T16:44:54 & 0.5 \\
\swift & 00032249051 & 2021-02-14T15:27:42 & 2021-02-14T20:36:53 & 1.5 \\
\swift & 00032249052 & 2021-02-21T14:53:45 & 2021-02-21T22:52:53 & 1.6 \\
\swift & 00032249053 & 2021-02-28T04:44:06 & 2021-02-28T20:26:52 & 1.3 \\
\swift & 00032249054 & 2021-03-07T10:06:21 & 2021-03-07T13:53:52 & 0.5 \\
\swift & 00032249055 & 2021-03-14T11:03:30 & 2021-03-14T19:15:51 & 1.5  \\
\swift & 00032249057 & 2021-03-23T13:44:47 & 2021-03-23T19:55:53 & 0.9  \\
\swift & 00032249058 & 2021-03-25T07:11:05 & 2021-03-25T13:43:52 & 1.0  \\
\swift & 00032249059 & 2021-03-28T10:05:35 & 2021-03-28T22:37:52 & 1.0  \\
\swift & 00032249060 & 2021-04-04T14:06:35 & 2021-04-04T14:12:52 & 0.4  \\
\swift & 00032249061 & 2021-04-07T02:27:37 & 2021-04-07T02:30:53 & 0.2  \\
\swift & 00032249062 & 2021-04-08T08:26:16 & 2021-04-08T13:16:53 & 0.5  \\
\swift & 00032249063 & 2021-04-11T00:13:19 & 2021-04-11T14:43:52 & 1.2  \\
\swift & 00032249064 & 2021-04-20T13:31:22 & 2021-04-20T17:17:52 & 1.3  \\
\swift & 00032249065 & 2021-04-25T20:57:04 & 2021-04-25T21:14:20 & 1.0  \\
\swift & 00032249066 & 2021-05-02T05:54:03 & 2021-05-02T06:04:52 & 0.6  \\
\swift & 00032249067 & 2021-05-04T04:22:09 & 2021-05-04T04:23:15 & 0.1  \\
\swift & 00032249068 & 2021-05-11T03:29:09 & 2021-05-11T03:41:54 & 0.8  \\
\swift & 00032249069 & 2021-05-16T03:10:55 & 2021-05-16T03:13:53 & 0.2  \\
\swift & 00032249070 & 2021-05-18T15:40:22 & 2021-05-18T15:42:52 & 0.1  \\
\swift & 00032249071 & 2021-05-23T18:12:35 & 2021-05-23T18:39:53 & 1.6  \\
\swift & 00032249072 & 2021-05-30T06:11:55 & 2021-05-30T06:38:52 & 1.6  \\
\swift & 00032249073 & 2021-06-06T19:51:30 & 2021-06-06T20:04:52 & 0.8  \\
\swift & 00032249074 & 2021-06-13T19:01:20 & 2021-06-13T19:29:52 & 1.7  \\
\swift & 00032249075 & 2021-06-20T10:39:31 & 2021-06-20T15:14:53 & 1.8  \\
\swift & 00032249076 & 2021-06-27T03:16:36 & 2021-06-28T10:03:52 & 2.1  \\
\swift & 00032249077 & 2021-07-04T07:29:20 & 2021-07-04T07:55:52 & 1.6  \\
\swift & 00032249078 & 2021-07-11T01:59:21 & 2021-07-11T02:26:53 & 1.6  \\
\swift & 00032249079 & 2021-07-18T07:36:20 & 2021-07-18T08:03:52 & 1.6  \\
\swift & 00032249080 & 2021-07-25T11:36:21 & 2021-07-25T12:01:50 & 1.5  \\
\swift & 00032249081 & 2021-08-01T15:37:16 & 2021-08-01T16:05:51 & 1.7  \\
\hline
\end{tabular}
}
\end{table}

\preto\tabular{\setcounter{rowcounter}{86}}
\newcounter{rowcounter}

\begin{table}
\caption*{Table~\ref{list_of_log}: continued}
\scalebox{0.65}{
\begin{tabular}{@{\makebox[1.8em][l]{\rownumber\space}} | llccr}
\hline
Instr. & Obs.ID & Start & Stop & Exp. \\
 & & \multicolumn{2}{c}{[YYYY-MM-DD hh:mm:ss (TT)]} & [ks]  
  \gdef\rownumber{\stepcounter{rowcounter}\arabic{rowcounter}} \\
\hline
\swift & 00032249082 & 2021-08-08T03:42:21 & 2021-08-08T04:10:54 & 1.7  \\
\swift & 00032249083 & 2021-10-29T13:57:23 & 2021-10-29T14:24:53 & 1.6  \\
\swift & 00032249084 & 2021-11-05T17:57:14 & 2021-11-05T18:24:52 & 1.7  \\
\swift & 00032249085 & 2021-11-12T09:13:15 & 2021-11-12T09:39:52 & 1.6  \\
\swift & 00032249086 & 2021-11-19T00:37:50 & 2021-11-19T18:35:53 & 2.8  \\
\swift & 00032249087 & 2021-11-26T01:21:13 & 2021-11-26T01:47:53 & 1.6  \\
\swift & 00032249088 & 2021-12-03T19:52:35 & 2021-12-03T20:19:52 & 1.6  \\
\swift & 00032249089 & 2021-12-10T17:23:04 & 2021-12-11T00:04:31 & 1.7  \\
\swift & 00032249090 & 2021-12-17T08:46:52 & 2021-12-17T09:14:52 & 1.7  \\
\swift & 00032249091 & 2021-12-24T06:13:28 & 2021-12-24T06:40:52 & 1.6  \\
\swift & 00032249092 & 2021-12-31T04:00:20 & 2021-12-31T04:27:52 & 1.6  \\
\swift & 00032249093 & 2022-01-07T04:37:21 & 2022-01-07T05:04:53 & 1.6  \\
\swift & 00032249094 & 2022-01-14T02:23:03 & 2022-01-14T02:51:53 & 1.7  \\
\swift & 00032249099 & 2022-02-18T07:55:51 & 2022-02-18T17:35:52 & 1.9  \\
\swift & 00032249100 & 2022-02-25T12:02:13 & 2022-02-26T04:05:40 & 2.1  \\
\swift & 00032249101 & 2022-03-04T14:23:45 & 2022-03-04T16:27:53 & 2.1  \\
\swift & 00032249102 & 2022-03-11T13:31:53 & 2022-03-11T21:53:53 & 2.2  \\
\swift & 00032249104 & 2022-03-18T16:04:26 & 2022-03-18T18:06:53 & 1.6  \\
\swift & 00032249105 & 2022-03-23T18:29:56 & 2022-03-23T18:42:53 & 0.8  \\
\swift & 00032249106 & 2022-03-25T07:06:13 & 2022-03-25T19:59:54 & 2.0  \\
\swift & 00032249107 & 2022-04-01T09:53:13 & 2022-04-01T16:26:53 & 2.3  \\
\swift & 00032249108 & 2022-04-08T19:42:45 & 2022-04-08T23:02:54 & 2.1  \\
\swift & 00032249109 & 2022-04-15T17:41:18 & 2022-04-15T22:40:52 & 2.1  \\
\swift & 00032249110 & 2022-04-22T11:39:30 & 2022-04-23T03:52:52 & 3.2  \\
\swift & 00032249111 & 2022-04-29T04:35:14 & 2022-04-29T04:47:54 & 0.8  \\
\swift & 00032249112 & 2022-05-04T16:39:51 & 2022-05-04T21:56:52 & 1.5  \\
\swift & 00032249113 & 2022-05-06T00:27:28 & 2022-05-06T16:53:52 & 2.2  \\
\swift & 00032249114 & 2022-05-13T07:44:31 & 2022-05-13T18:56:52 & 2.6  \\
\swift & 00032249115 & 2022-05-20T10:10:09 & 2022-05-20T11:50:52 & 1.0  \\
\swift & 00032249116 & 2022-05-27T04:22:29 & 2022-05-27T09:08:53 & 2.4  \\
\swift & 00032249117 & 2022-05-29T11:51:34 & 2022-05-29T13:32:22 & 1.7  \\
\swift & 00032249118 & 2022-06-01T15:09:12 & 2022-06-01T15:16:54 & 0.5  \\
\swift & 00032249119 & 2022-06-03T05:09:19 & 2022-06-03T06:53:52 & 2.0  \\
\swift & 00032249120 & 2022-06-10T05:41:12 & 2022-06-10T06:01:53 & 1.2  \\
\xmm & 0883960201 & 2022-06-13T20:42:51 & 2022-06-15T06:59:59 & 130 \\
\swift & 00032249121 & 2022-06-17T11:12:18 & 2022-06-17T14:48:53 & 1.7  \\
\swift & 00032249122 & 2022-06-24T07:04:44 & 2022-06-24T10:42:51 & 2.0  \\
\swift & 00032249124 & 2022-07-08T02:04:53 & 2022-07-08T08:40:53 & 2.5  \\
\swift & 00032249125 & 2022-07-15T15:37:13 & 2022-07-15T18:57:52 & 2.3  \\
\swift & 00032249126 & 2022-07-22T06:35:09 & 2022-07-22T11:35:52 & 2.5  \\
\swift & 00032249127 & 2022-07-29T05:39:56 & 2022-07-29T06:05:53 & 1.6  \\
\swift & 00032249128 & 2022-08-05T03:17:23 & 2022-08-06T20:33:52 & 2.3  \\
\swift & 00032249129 & 2022-10-29T04:34:32 & 2022-10-29T08:12:54 & 2.0  \\
\swift & 00032249130 & 2022-11-04T06:54:36 & 2022-11-04T22:44:52 & 2.1  \\
\swift & 00032249131 & 2022-11-11T04:31:52 & 2022-11-11T18:59:52 & 2.0  \\
\swift & 00032249132 & 2023-01-21T17:26:50 & 2023-01-21T22:39:53 & 2.0  \\
\swift & 00032249133 & 2023-01-28T14:43:42 & 2023-01-28T16:46:52 & 0.8  \\
\swift & 00032249134 & 2023-02-04T09:00:02 & 2023-02-04T11:01:52 & 2.3  \\
\swift & 00032249135 & 2023-02-11T10:52:32 & 2023-02-11T14:28:52 & 2.1  \\
\swift & 00032249136 & 2023-02-18T20:50:10 & 2023-02-18T22:50:52 & 2.0  \\
\swift & 00032249137 & 2023-02-25T19:48:48 & 2023-02-25T21:51:53 & 2.1  \\
\swift & 00032249138 & 2023-03-04T10:40:55 & 2023-03-04T22:06:53 & 1.7  \\
\swift & 00032249139 & 2023-03-11T12:46:49 & 2023-03-11T23:59:54 & 1.2  \\
\swift & 00032249140 & 2023-03-15T11:56:14 & 2023-03-15T12:10:54 & 0.9  \\
\swift & 00032249141 & 2023-03-18T16:09:43 & 2023-03-18T17:53:52 & 2.1  \\
\swift & 00032249142 & 2023-03-25T16:58:10 & 2023-03-25T20:18:52 & 2.1  \\
\swift & 00032249143 & 2023-04-01T02:33:36 & 2023-04-01T09:13:52 & 2.2  \\
\swift & 00032249144 & 2023-04-08T03:02:08 & 2023-04-08T10:54:52 & 2.4  \\
\swift & 00032249145 & 2023-04-15T03:07:04 & 2023-04-15T16:01:54 & 1.2  \\
\swift & 00032249146 & 2023-04-19T10:49:39 & 2023-04-19T21:59:52 & 1.1  \\
\swift & 00032249147 & 2023-04-22T11:34:13 & 2023-04-22T11:58:53 & 1.5  \\
\swift & 00032249148 & 2023-04-26T12:15:35 & 2023-04-26T12:28:53 & 0.8  \\
\swift & 00032249149 & 2023-04-28T21:44:42 & 2023-04-29T07:33:53 & 0.9  \\
\swift & 00032249150 & 2023-05-03T04:41:26 & 2023-05-03T09:59:53 & 1.4  \\
\swift & 00032249151 & 2023-05-06T10:25:22 & 2023-05-06T16:55:54 & 2.2  \\
\swift & 00032249152 & 2023-05-13T12:13:00 & 2023-05-13T12:39:52 & 1.6  \\
\swift & 00032249153 & 2023-05-17T05:28:22 & 2023-05-17T05:43:53 & 0.9  \\
\swift & 00032249154 & 2023-05-20T09:16:51 & 2023-05-20T11:00:52 & 2.2 \\
\swift & 00032249155 & 2024-01-15T03:37:12 & 2024-01-15T21:09:53 & 2.4  \\
\swift & 00032249156 & 2024-01-29T09:13:06 & 2024-01-29T23:47:52 & 2.8  \\
\swift & 00032249157 & 2024-02-12T13:19:57 & 2024-02-12T21:31:53 & 2.6  \\
\swift & 00032249158 & 2024-02-26T15:42:23 & 2024-02-26T17:32:53 & 2.6  \\
\swift & 00032249159 & 2024-03-11T11:58:38 & 2024-03-11T21:40:53 & 2.3  \\
\swift & 00032249161 & 2024-04-08T14:49:49 & 2024-04-08T16:51:53 & 0.9  \\
\swift & 00032249162 & 2024-04-11T07:32:56 & 2024-04-11T07:53:52 & 1.3  \\
\swift & 00032249163 & 2024-04-22T10:48:30 & 2024-04-22T14:22:54 & 2.6  \\
\swift & 00032249164 & 2024-05-06T11:10:25 & 2024-05-06T13:13:53 & 1.1 \\ 
\swift & 00032249165 & 2024-05-20T05:05:03 & 2024-05-20T08:39:52 & 2,5  \\
\swift & 00032249167 & 2024-06-03T14:55:43 & 2024-06-03T15:22:52 & 1.6  \\
\swift & 00032249168 & 2024-06-06T02:48:31 & 2024-06-06T03:09:53 & 1.3  \\
\swift & 00032249169 & 2024-06-17T07:19:18 & 2024-06-17T12:27:51 & 2.5  \\
\swift & 00032249170 & 2024-07-01T05:45:05 & 2024-07-01T10:45:53 & 2.8  \\
\swift & 00032249171 & 2024-07-15T04:08:48 & 2024-07-15T07:46:52 & 3.1  \\
\swift & 00032249172 & 2024-07-29T07:19:41 & 2024-07-29T12:29:51 & 2.7  \\
\hline
\end{tabular}
}
\end{table}

\label{lastpage}

\end{document}